\documentclass[pdflatex,sn-mathphys-num]{sn-jnl}

\usepackage{graphicx}%
\usepackage{multirow}%
\usepackage{amsmath,amssymb,amsfonts}%
\usepackage{amsthm}%
\usepackage{mathrsfs}%
\usepackage[title]{appendix}%
\usepackage{xcolor}%
\usepackage{textcomp}%
\usepackage{manyfoot}%
\usepackage{booktabs}%
\usepackage{algorithm}%
\usepackage{algorithmicx}%
\usepackage{algpseudocode}%
\usepackage{listings}%

\usepackage{tabularx}
\usepackage{booktabs}
\usepackage{multirow}
\usepackage{subcaption}
\usepackage{placeins}
\usepackage{lineno}

\theoremstyle{thmstyleone}

\theoremstyle{thmstyletwo}%

\theoremstyle{thmstylethree}%

\begin{document}

\title{The efficiency-gain illusion: People underestimate the rate of AI use and overestimate its benefits on simple tasks}

\author*[1]{\fnm{Sunny} \sur{Yu}}\email{syu03@stanford.edu}

\author[1]{\fnm{Myra} \sur{Cheng}}\email{myra@cs.stanford.edu}

\author[1]{\fnm{Ahmad} \sur{Jabbar}}\email{jabbar@stanford.edu}

\author[2]{\fnm{Ilia} \sur{Sucholutsky}}\email{is3060@nyu.edu}

\author[3, 4]{\fnm{Katherine M.} \sur{Collins}}\email{katiemc@mit.edu}

\author[1]{\fnm{Dan} \sur{Jurafsky}}\email{jurafsky@stanford.edu}

\author[1]{\fnm{Robert D.} \sur{Hawkins}}\email{rdhawkins@stanford.edu}

\affil[1]{\orgname{Stanford University}}
\affil[2]{\orgname{New York University}}
\affil[3]{\orgname{MIT}}
\affil[4]{\orgname{Princeton AI Lab}}

\abstract{

People are increasingly turning to AI assistance for simple tasks, e.g., arithmetic, spell-check, and answering simple questions. But does AI assistance actually save users time and effort? We investigate people's propensity to use AI for cognitively simple tasks and assess whether their reliance is well-calibrated. Across three pre-registered user studies (N = 2691), we find that people frequently choose to use AI even when doing so is inefficient (i.e. provides no meaningful time or effort savings). We identify systematic miscalibration at two levels: (1) a self-estimate miscalibration where people on average believe that they are using AI less than they actually are, and (2) efficiency-gain illusions where people overestimate how much time and effort savings AI use affords. We also identify a session-level carryover effect where a participant's prior AI use leads to further AI adoption and entrenches their miscalibration about time savings. Our results shed light on the mechanisms and  biases underlying people's choice of whether to use AI as well as the risk of an overreliance feedback loop.

}


\maketitle

\section{Introduction}\label{sec1}

People increasingly use large language models (LLMs) for a wide range of tasks, from coding and calculations to writing assistance to personal advice \citep{collins2024building,zao2025people, cheng2026sycophantic}. A common narrative is that LLM-based AI chatbots can dramatically boost productivity and efficiency \citep{handa2025economic, wang2025ai, tamkinmccrory2025productivity, anthropic2026aeiv4,xiong2024search, hooper2025cognitive}. However, there is emerging evidence that assistance from artificial intelligence (AI) does not necessarily save time or effort \citep{stadler2024cognitive,dhillon2024shaping}. For example, \citet{becker2025measuring} found that contrary to expectations, AI assistance slowed down professional software developers' coding time by 19\%. 

The decision of whether to seek external assistance on a given task depends on a cost-benefit comparison: people should delegate to external tools when internal costs exceed the costs of using external support \citep{risko2016cognitive}. Central to this comparison is \emph{calibration}, i.e., having an accurate mental model of both one's own capabilities and those of the external tool. If AI does not provide efficiency gains on simple tasks that can be completed quickly without AI, such as basic arithmetic, editing a sentence, or summarizing a paragraph, then the decision to use AI for them would be inefficient. In this study, we investigate the mechanisms underlying people's engagement with LLM chatbots and seek to understand whether using AI on such tasks is counterproductive from an efficiency-gain perspective.

We conducted three large-scale pre-registered studies ($N=2691$) to understand people's decision to use AI assistance (defined as prompting an LLM-based AI chatbot in this study) and the mechanisms behind this decision. In Study 1, we explore whether people accurately estimate how often they would rely on AI and whether AI use saves time and effort. Comparing the population-level predicted (where participants chose whether they would complete a task independently or using AI) versus actual rates of AI use, we identified a \textbf{self-estimate miscalibration}: people underestimate their self-reported proclivity towards using AI. Furthermore, using AI did not lead to efficiency gains.

In Study 2, we seek to understand why people might use AI for simple tasks despite AI use not providing efficiency benefits. One hypothesis is that people are miscalibrated about how much time and effort AI assistance saves. To test this hypothesis, we compared people's predicted versus actual time and effort completing these tasks with and without AI assistance and identified  \textbf{efficiency-gain illusions}, where people overestimated both the time and effort that AI saves. On average, people predicted AI assistance to save time by 55.7 seconds when it only saved 7.5 seconds. This miscalibration is particularly severe on the simple variants of the tasks, where people predicted AI assistance to save time, but it made tasks slower to complete in reality. 

Finally, in Study 3, we investigate whether using AI improves or worsens people's calibration about AI's efficiency gains by randomly assigning participants to different completion modes (with AI vs. independently) and measuring how these completion modes affect subsequent AI adoption. We found that after people used AI, they became more miscalibrated about how much time AI saves and exhibited a higher rate of subsequent AI use, which may exacerbate this miscalibration and pose risks of an overreliance feedback loop. 

A high rate of AI use, especially on simple tasks, poses risks such as cognitive deskilling \citep{gerlich2025ai}, reduced confidence \citep{lee2025impact}, and impairments in independent performance \citep{liu2026ai}, knowledge retention \citep{barcaui2025chatgpt} and learning abilities \citep{shen2026ai}. Together, our findings provide evidence of people's cognitive biases about their interactions with AI systems: average predicted AI use rates underestimate observed use, and average predicted time and effort savings exceed observed gains. These miscalibrations may in turn exacerbate inefficient use of AI and form a feedback loop that perpetuates miscalibration and overreliance.

\FloatBarrier

\begin{figure*}[ht] 
    \centering
    \includegraphics[width=0.99\textwidth]{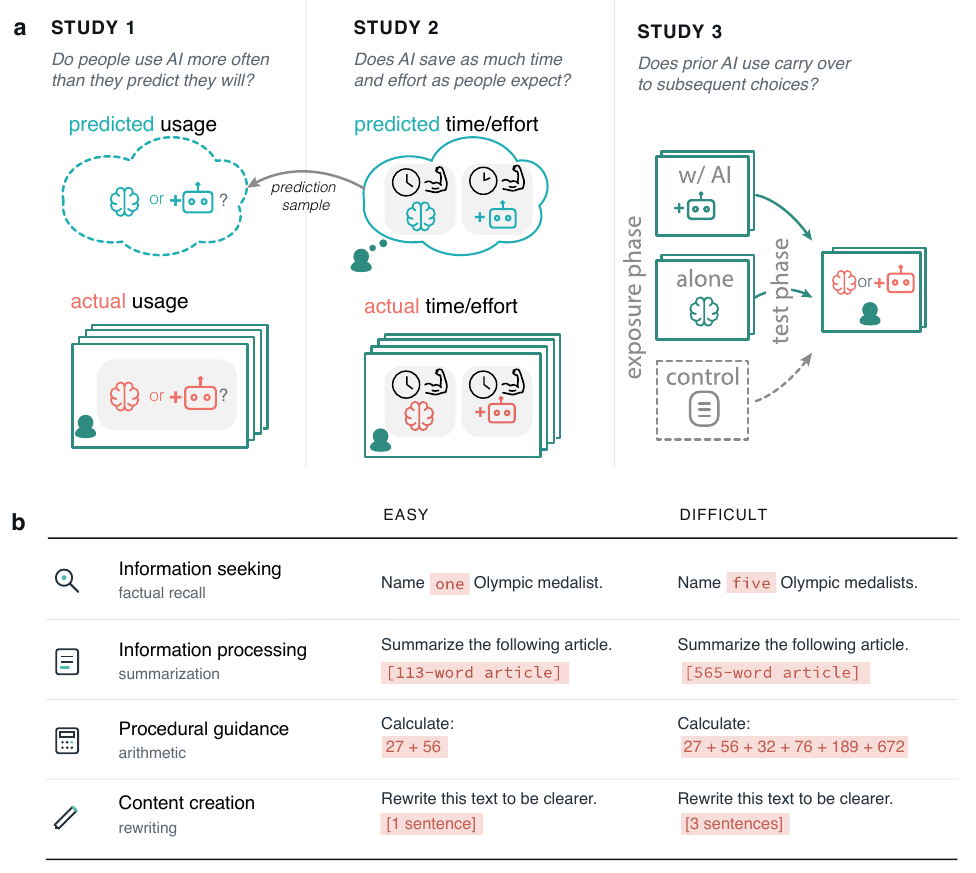} 
    \caption{(a). \textbf{Overview of the study design.} In Study 1 we compared the predicted rate of AI use with the actual rate of AI use. In Study 2, to measure efficiency-gain illusions, we compared predicted time and effort with actual time and effort from participants who completed the same tasks with or without AI assistance.
    In Study 3, to understand whether using AI improves calibration, participants were exposed to using AI or completing tasks independently in the exposure phase, and we measured the subsequent AI adoption rate. (b) \textbf{Examples of simple tasks.} The tasks that we studied are drawn from a taxonomy of how people use AI, spanning four distinct categories with three task types per category and  an easy variant and a difficult variant per task. }

    \label{fig:fig1}
\end{figure*}
\FloatBarrier

\section{Results}\label{sec2}
We study AI use on simple tasks by constructing a set of 24 tasks derived from the Taxonomy of User Needs and Actions (TUNA), a framework of different ways that people use AI based on different degrees of cognitive effort required \citep{shelby2025taxonomy}. Our tasks capture TUNA's four cognitively distinct categories: information seeking, information processing and synthesis, procedural guidance and execution, and content creation and transformation (see Table \ref{tab:tasks} for full details). In the taxonomy, each category has 3 task types; for each task type, we constructed two variants, either relatively easier or more difficult (e.g., ``Name one Olympic medalist.'' vs. ``Name five Olympic medalists.'') Though we will henceforth distinguish between these variants as ``easy'' versus ``difficult'' tasks, note that all tasks are relatively simple and can be completed in under five minutes independently (see Appendix \ref{appendix:task_details}). 

In Study 1, we sought to understand whether people use AI on simple tasks and whether they do so more than the population-level expected rate. Participants completed four tasks and could choose whether to use an LLM-based AI chatbot for assistance on each task. When using the chatbot, participants were allowed to prompt it however they wished. We compared this behavioral rate with a predicted rate, obtained from a separate sample of participants who stated whether they would prefer to complete a task independently or with AI assistance. Note that here, we use a between-subjects design to avoid participants' behaviors and decisions being affected by their stated preference as a result of cognitive dissonance \citep{festinger1962cognitive}. Our between-subject design instead compares the population-level predicted self-estimate with the behavioral average. While this means that we cannot directly examine specific individuals' metacognitive errors regarding their choice to use AI, it nonetheless enables an assessment of whether people  use AI more than the self-reported prediction rates.

In Study 2, we compared predicted time and effort (based on subjective perception using the NASA-TLX scale \citep{Hart_1986}) versus actual time and effort. In the prediction sample, participants predicted the time and effort required for independent and AI-assisted completions; in the completion sample, participants were randomly assigned to completing four tasks independently or using AI, during which we recorded their time and effort. The study design is again between-subject; otherwise, the actual completion times and effort would be biased by participants' familiarity with the tasks. 

In Study 3, we investigated whether using AI leads to subsequent choices to use AI. Participants first either completed two tasks independently or using AI in an ``exposure'' phase, then we recorded whether participants chose to use AI for each of the two easy task variants in a ``test'' phase. Full methods details are in Section \ref{sec:methods}.

\subsection{People do not realize how much they use AI on simple tasks}

\begin{figure}[t] 
    \centering
    \includegraphics[width=0.8\columnwidth]{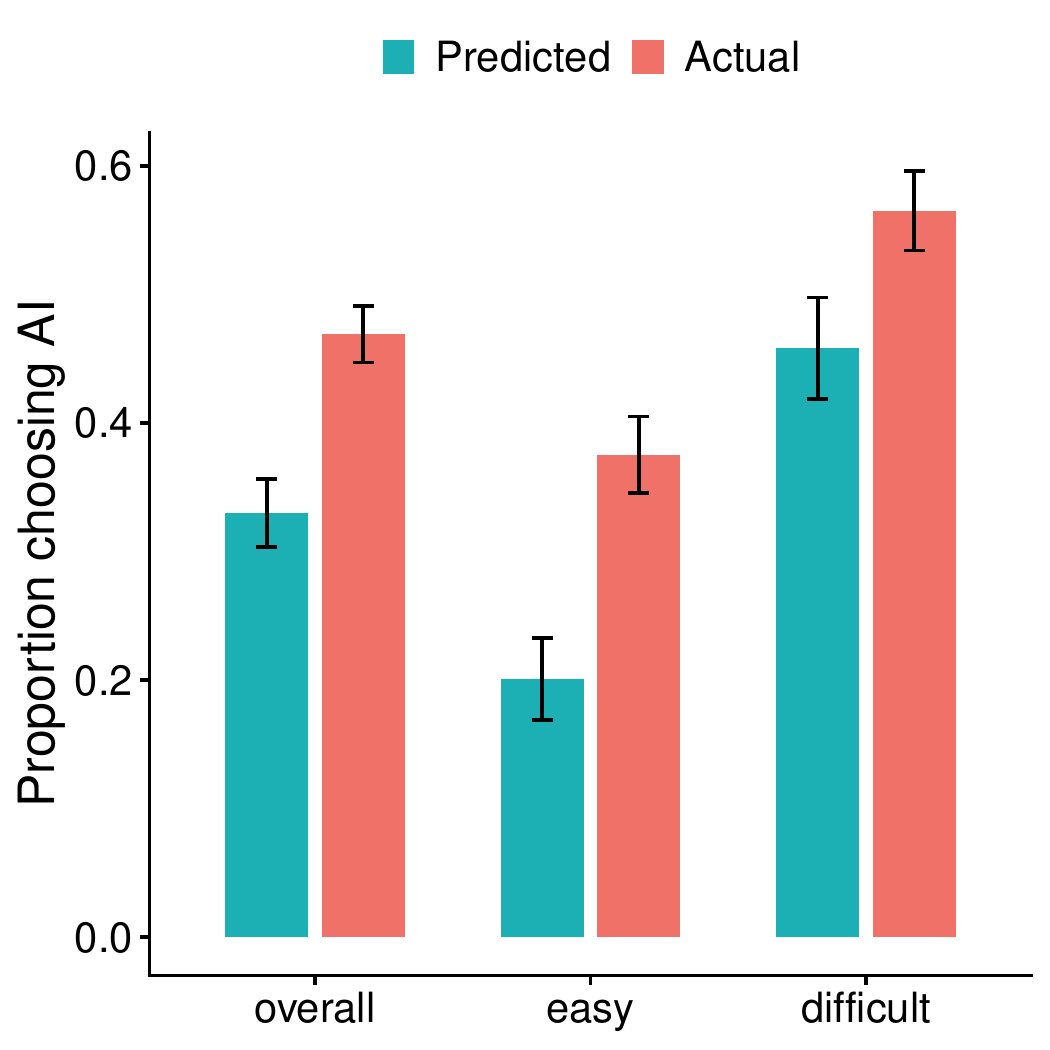} 
    \caption{\textbf{People use AI significantly more than the predicted rate for both easy and difficult task variants.} The population-level rate of AI use (red bars) is significantly greater than the predicted rate of AI use (blue bars). Specifically, the overall rate of actual AI use is 47\%, which is 14 percentage points greater than the predicted rate (33\%). The difference is larger on easy task variants, where there is an 18\% difference between the 20\% predicted rate and the 38\% actual rate of AI use. Bars show the mean proportion of using AI with 95\% CIs (1.96 ± SE). All $p<0.001$. }
    \label{fig:fig3}
\end{figure}

Metacognitive monitoring provides us with imperfect but usable signals about our own cognitive processes, like how long a task will take, how difficult it feels, and whether we are on track \citep{koriat2015metacognition, yeung2012metacognition, fleming2017self,dubey2021aha}. We are interested in understanding whether there are problems in metacognitive monitoring for AI use, and in particular whether the rate at which people predict they will use AI matches the actual behavioral rate. To test whether people exhibit a \textbf{self-estimate miscalibration} of relying on AI more than the population-level anticipated average, we asked participants to complete four tasks with the option to use AI assistance on each task. We contrast this rate to a stated preference rate, obtained from participants stating whether they would personally prefer to complete tasks independently versus with AI assistance (\textit{``If you need to complete the task, how would you prefer to complete it? A: Independently B: With the help of an AI model''}). 

We find that people from Study 1 indeed used AI significantly more than the average predicted rate. On average, participants reported that they would use AI in 33\% of tasks, but the population-level rate of AI use is 47\% ($\beta = 1.07$, $p < 0.001$). This gap is larger for easy task variants ($\beta=0.69$, $p<0.001$): participants predicted a 20\% AI use but the actual rate of AI use was 38\% ($\beta = 1.42$, $p < 0.001$), nearly doubling the stated preference rate. 

On difficult task variants, the gap is smaller (46\% predicted vs. 57\% actual; $\beta=0.73$, $p<0.001$), indicating that the self-estimate miscalibration is disproportionately higher for easy task variants. Despite this high rate of AI use, we find no evidence of time or effort savings from using AI vs. completing tasks independently: for time, there is no significant difference  ($\beta=-5.6$, $p=0.10$); for effort, people who completed tasks independently actually reported lower mental effort (1.99 for independent vs. 2.11 for AI-assisted on a 7-point scale) than those who used AI ($\beta=-0.12$, $p<0.01$). These findings are robust to controlling for individual participant traits such as Need For Cognition (Appendix \ref{appendix:study1_stats}) \citep{cacioppo1982need}, hold across categories and tasks (Appendix \ref{appendix:study1_categories}), and are consistent after removing extreme data (i.e. participants who used AI on all the tasks or never used AI; Appendix \ref{appendix:study1_behavior}).

Notably, for a wide range of tasks, while virtually no participants (the average predicted rate is 2.3\% across five of the tasks) predicted that they would use AI, in the actual completion sample, 21-54\% participants chose to use AI for the task ($\beta=0.25$, $p<0.01$; $\beta=0.32$, $p<0.001$), a rate significantly higher than the predicted one. These tasks included information seeking tasks like simple color recall (``Name the color that you get when you mix white and black''; $\beta=0.35$, $p<0.001$), procedural execution tasks like arithmetic (``What is 27+56''; $\beta=0.23$, $p<0.01$), and tasks that required information processing, e.g. the trolley problem ($\beta=0.18$, $p<0.05$). The findings demonstrate that people not only use AI at a high rate on simple task variants but do so at a rate significantly higher than what people would predict for themselves. In the next section, we explore underlying mechanisms behind the high rate of usage and the predicted and behavioral mismatch.

\begin{figure}[t]
    \centering
      \includegraphics[width=0.9\columnwidth]{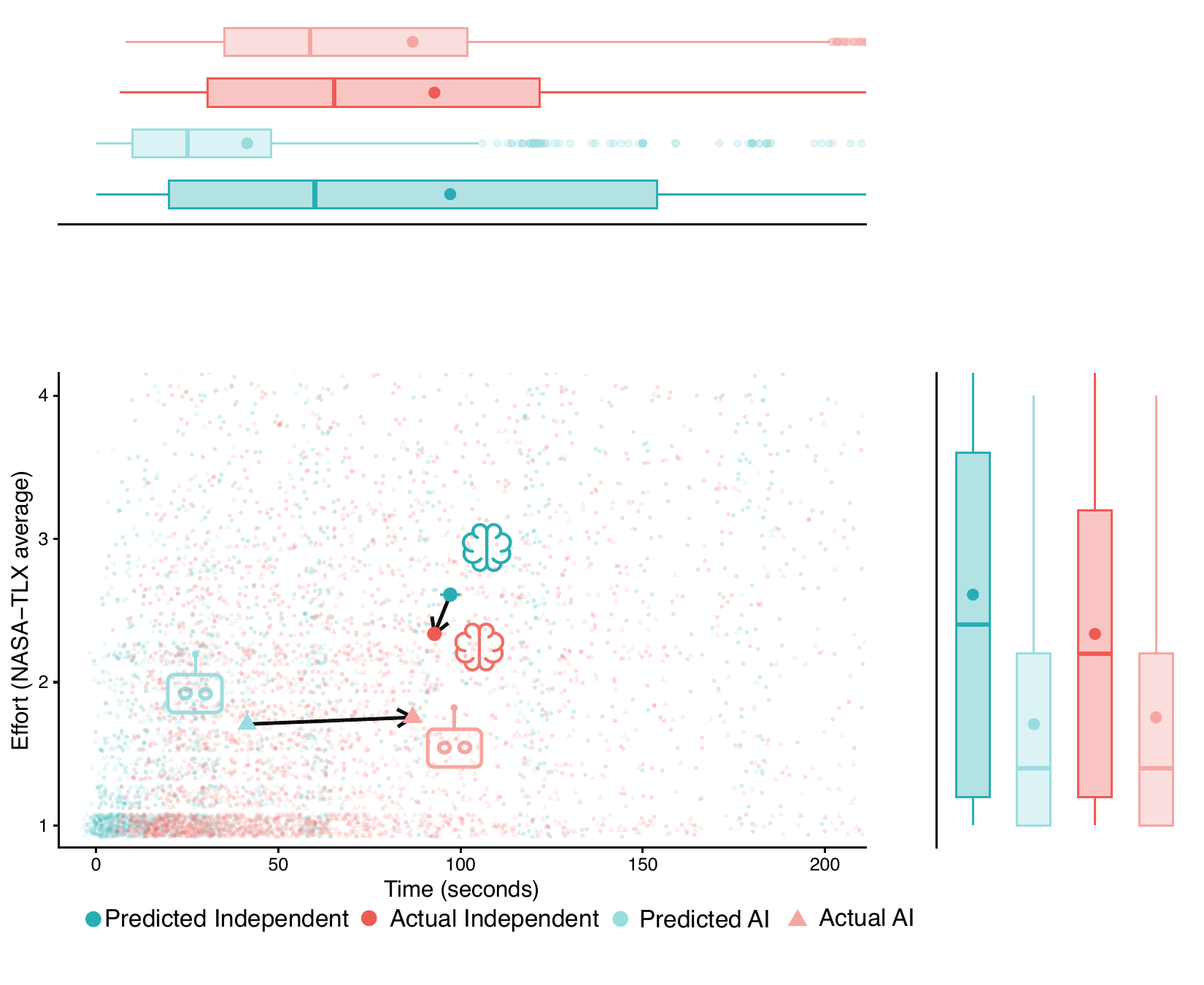} 
    \caption{\textbf{Predicted (blue) vs. actual (red) time and effort for the independent vs. AI-assisted conditions.} The main plot shows how the actual time or effort deviates from the predicted ones for each condition. The dot is the mean value (predicted AI-assisted, actual AI-assisted, actual independent, and predicted independent; from left to right), and the arrow shows the direction of difference between the predicted and actual ones. For the AI condition, we identify a \textbf{speedup illusion} where the actual time (86.2 seconds) is significantly greater than the predicted one (43.3 seconds); no significant difference was found between the actual vs. predicted effort. For predictions about independent completion, the opposite pattern emerges: participants are well-calibrated about the completion time, with no significant difference between the predicted vs. actual time (99 seconds vs. 93.7 seconds), while participants overestimated the effort required (2.66 predicted vs. 2.36 actual). The top marginal plot shows the box plot for the x-axis (time), and the right marginal plot shows the box-plot for the y-axis (NASA-TLX). The line represents the median value, and the dot represents the mean. }
     \label{fig:figure3}
\end{figure}

\subsection{On average, people overestimate how much time and effort AI assistance saves.}

People routinely offload cognition to external tools (e.g., writing, calculators, others, web search) \citep{leimeister2010collective,fan2023drawing}. From a resource-rational perspective, this reflects adaptive allocation of limited cognitive resources \citep{lieder2020resource,griffiths2019doing,griffiths2020understanding}. People delegate when internal costs exceed those of external support, especially under high difficulty or cognitive load \citep{dunn2016toward,wahn2023offloading}. Thus, we hypothesize that the high rate of AI use might be associated with people's general expectations of efficiency gains. We further hypothesize that people overestimate these efficiency gains.

We find that people generally expect AI assistance to reduce both task completion time and subjective mental effort. On average, participants predicted AI assistance to reduce completion time by 58\%, from 97.1 seconds to 41.4 seconds ($p<0.001$). Similarly, they predicted AI assistance to reduce mental effort by 35\%, from 2.61 to 1.71 on the 7-point NASA-TLX scale ($p<0.001$), which measures people's self-reported subjective effort (more details in Appendix \ref{appendix:nasa}). This expectation for an efficiency gain is not a generic delegation miscalibration but is AI-specific: we used ``another participant'' as a baseline in the prediction sample and found that even though participants predicted assistance from another participant to reduce time (from 94 seconds to 83.2 seconds; $p<0.001$), the difference is significantly smaller in magnitude than the predicted difference when the assistance source is AI ($\beta=-44.9$, $p<0.001$). Similarly, participants predicted an effort reduction of 0.12 points when assistance comes from another participant ($p<0.05$), but the magnitude is significantly smaller than what people predicted for AI assistance ($\beta=-0.78$, $p<0.001$). Therefore, participants expect offloading to AI to be even more efficient than offloading to another participant.

By comparing participants' predictions to actual completion times and effort ratings, we identify two \textbf{efficiency-gain illusions}. First, we find that participants overestimated how much time AI assistance saves.
Formally, for a task $\tau$, let $t_H(\tau)$ denote the mean time for humans to complete the task independently, and let $t_A(\tau)$ denote the mean completion time with AI assistance. Let $\hat{t}_H(\tau)$ and $\hat{t}_A(\tau)$ denote the corresponding average predicted completion times. We find a \textbf{speedup illusion} where \begin{equation}
\hat{t}_H(\tau) - \hat{t}_A(\tau) > t_H(\tau) - t_A(\tau).
\label{equation:time_delta}
\end{equation} 
Second, we find that people overestimated how much cognitive effort AI assistance alleviates. Let $E_H(\tau)$ and $E_A(\tau)$ similarly represent the actual subjective mental effort reported after completing the task independently and with AI assistance, respectively, and let $\hat{E}_H(\tau)$ and $\hat{E}_A(\tau)$ denote the corresponding predicted effort ratings. The \textbf{offloading illusion} occurs when the predicted effort reduction from AI assistance exceeds the actual effort reduction:
\begin{equation}
\hat{E}_H(\tau) - \hat{E}_A(\tau) > E_H(\tau) - E_A(\tau).
\label{equation:effort_delta}
\end{equation}

While we find the same overall illusion for both time and effort, we identify distinct patterns (Figure \ref{fig:figure3}). For time, participants have a distorted sense of how much \textit{AI} speeds up the task, i.e., $\hat{t}_A(\tau) > t_A(\tau)$, while they are able to accurately predict the time spent to complete the task independently, i.e., $\hat{t}_H(\tau) \approx {t}_H(\tau)$. On the other hand, for effort, participants generally imagine the task to be harder to complete independently, i.e., $\hat{E}_H(\tau) > E_H(\tau)$, whereas they are able to accurately assess the (relatively low) effort expended when using AI, i.e., $\hat{E}_A(\tau) \approx {E}_A(\tau)$.

Specifically, participants expected AI to save more time (55.7 seconds) than it actually did (7.5 seconds) (Equation~\eqref{equation:time_delta}; $\beta=-48.33$, $p<0.001$). This difference primarily arises from miscalibration about AI-assisted performance, i.e., $\hat{t}_A(\tau)$ versus $t_A(\tau)$: the actual completion time with AI (86.2 seconds) is nearly double the predicted time with AI (43.3 seconds) ($\beta=42.9$, $p<0.001$).
The speedup illusion held across all four task categories and both difficulty levels (Appendix \ref{appendix:study2_variations}). Notably, 
the speedup illusion is unique to AI assistance. There was no significant difference between the predicted (99 seconds) and actual (93.7 seconds) completion times in the independent completion condition ($\beta=-5.38$, $p=0.08$), which suggests that people are well-calibrated about their own performance. Moreover, we find that NFC positively predicts $\hat{t}_H(\tau) - \hat{t}_A(\tau)$, the difference between the predicted time for independent vs. AI-assisted completions, meaning that people who are less willing to think are even more susceptible to the speedup illusion  (Appendix \ref{appendix:study2_traits}).

For the offloading illusion, we found that while
participants accurately estimated the amount of effort $E_A$ required to complete a task with AI assistance (predicted: 1.76 vs. actual: 1.74; $\beta=-0.02$, $p=0.63$), they overestimated $E_H$, the amount of effort required for them to complete the task independently (predicted: 2.66 vs. actual: 2.36; $\beta=-0.30$, $p<0.001$). As a result, the magnitude of effort savings is overestimated, with an average predicted reduction of 0.90 compared to an actual reduction of 0.62 (Equation~\eqref{equation:effort_delta}; $\beta=-0.28$, $p<0.001$). The findings generalize to most individual items in the NASA-TLX scale (Appendix \ref{appendix:study2_nasa_items}) and are robust against the prediction order (predicting time before effort vs. predicting effort before time; Appendix \ref{appendix:prediction_order}), revealing a systematic bias toward perceived efficiency gains: people overestimate how much \textbf{time} and \textbf{effort} AI saves on cognitively simple tasks.

Across all tasks, AI assistance did not significantly reduce completion time overall ($\beta=6.17$, $p=0.07$). A decomposition of the AI-assisted completion time reveals that this is because current chat-interface friction dominates on trivial tasks, leading to a slow-down effect where AI-assisted completion times is longer than the independent completion by 10.0 seconds on easy task variants (from 60.2 seconds for independent to 70.2 seconds for AI-assisted; $p<0.05$). We decomposed the completion time into three components --- the time it takes to construct the prompt, the time for the model to generate a response, and the time the user reads and processes the model response, i.e., $t_A(\tau) = t_P(\tau) +t_M(\tau) + t_R(\tau)$ --- and found that across all tasks, prompting (48.7 seconds) took significantly longer than response processing (37.6 seconds) ($\beta=11.1$, $p<0.001$). Moreover, 41\% of the prompts were directly copied and pasted from the instructions, showing that on easy task variants, copying and pasting prompts took longer than completing the tasks independently, yet participants still chose to use AI at a high rate. The friction highlights how AI use can be inefficient as the speedup illusion persists.

\begin{figure}[t]
    \centering
    
    \begin{subfigure}[t]{0.48\columnwidth}
        \centering
        \includegraphics[width=\linewidth]{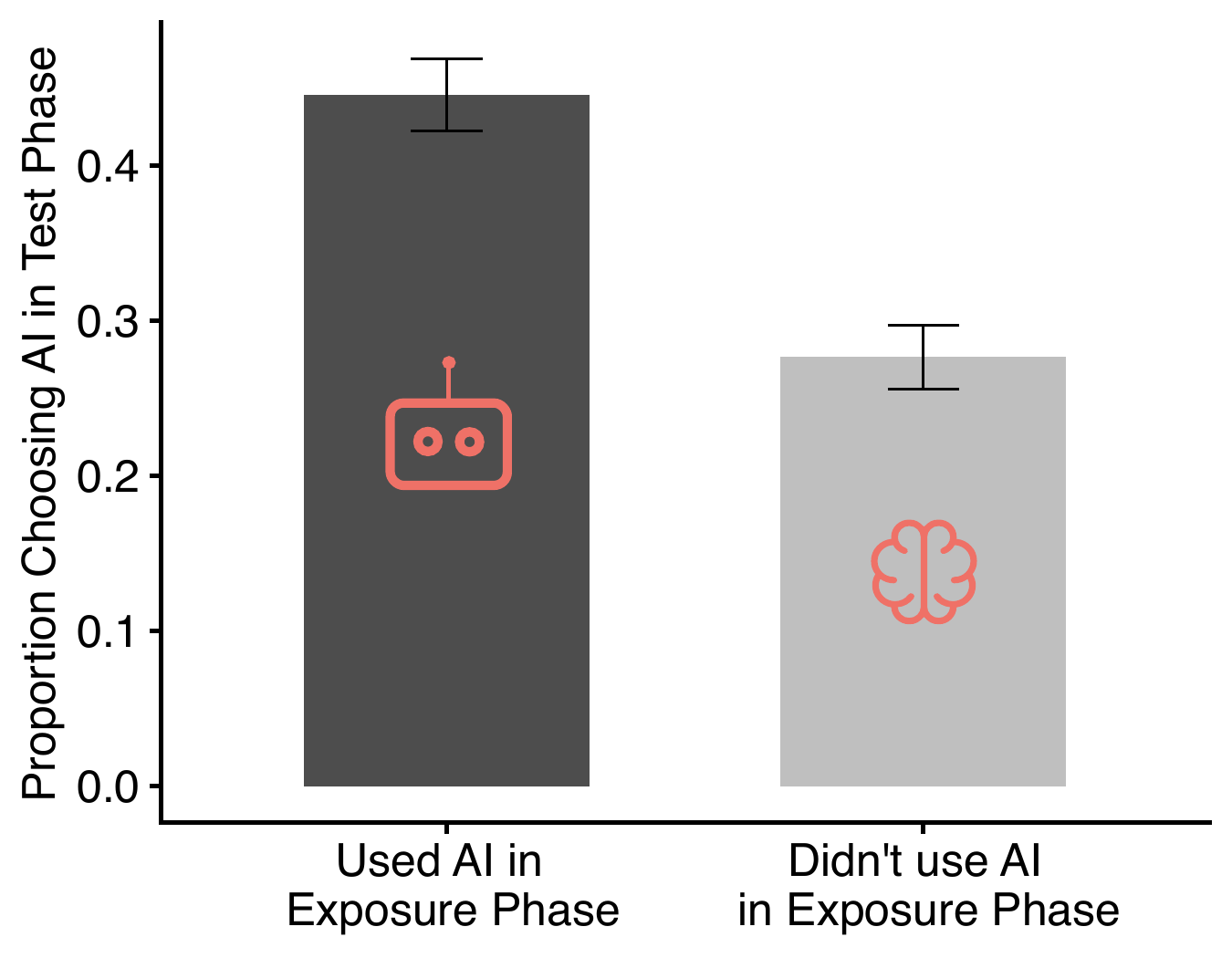}
    \end{subfigure}
    \hfill
    \begin{subfigure}[t]{0.48\columnwidth}
        \centering
        \includegraphics[width=\linewidth]{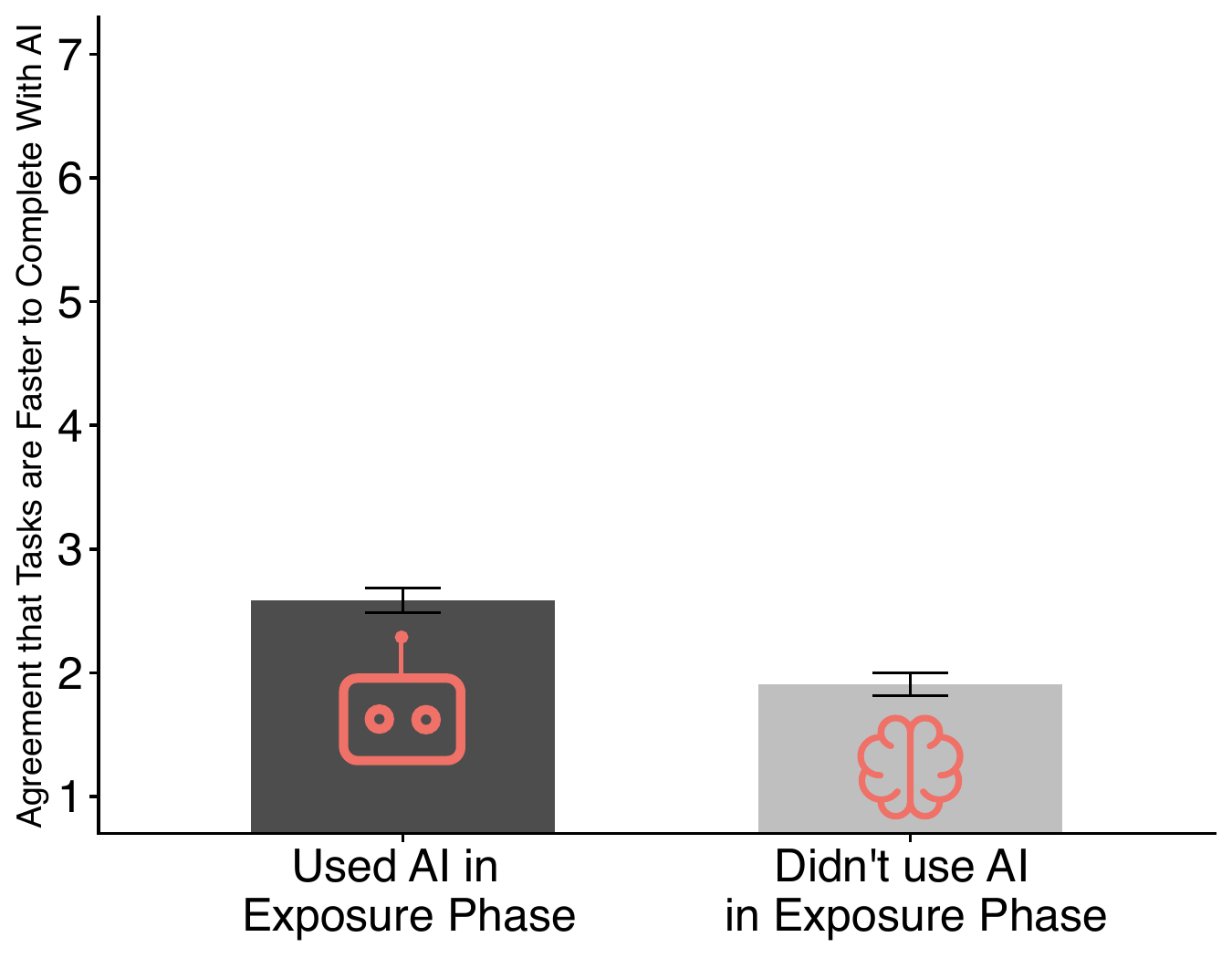}
    \end{subfigure}
    
    \caption{\textbf{AI use in the exposure phase leads to a higher rate of AI use and a higher agreement with the speedup illusion compared to people who completed tasks independently in the exposure phase.} 
   Left: \textbf{AI use rate by completion mode in exposure phrase}: participants who completed the first two tasks with AI were more likely to use AI again on the trials in the test phase (44.5\% vs. 27.7\%). Right: \textbf{AI exposure leads to a higher agreement with the speedup illusion}: Participants who were previously exposed to completing a task with AI reported lower agreement with the statement that tasks in the test phase were faster to complete independently. Participants provided ratings on a 7-point scale, and we reverse coded the mean value so that higher means greater agreement with the speedup illusion. Error bars capture 95\% CI; all $p<0.001$.}
    \label{fig:study3_main}
\end{figure}

\subsection{AI use encourages further AI use and exacerbates miscalibration.}

Since LLMs are an emerging technology, it is possible that people inaccurately predict efficiency gains because they are not familiar with AI. Here, we study whether using AI makes people better calibrated about its efficiency gains. In Study 3, we first assigned participants to complete two tasks either independently or with AI assistance in an ``AI exposure'' phase; we also varied whether the two tasks they completed in this phase were an easy or difficult variant. Then, we assessed how this exposure affected their subsequent choice to use AI on two easy task variants and their assessments about whether the tasks would be faster to complete independently vs. AI.

Contrary to the possibility that experience improves calibration, we identified a \textbf{session-level carryover effect} where initial AI use increases subsequent AI use. Participants who initially completed tasks with AI became even more likely to opt for AI assistance on easy task variants, even though doing so did not offer time or effort savings on average. Participants who completed tasks with AI during the exposure phase chose AI assistance on 44.5\% of subsequent tasks, compared to 27.7\% among participants who completed the exposure tasks independently ($\beta=0.54$, $p<0.001$). This effect held regardless of whether participants were exposed to completing either easy or difficult task variants in the exposure phase and are robust against NFC (Appendix \ref{appendix:study3_stats}) and holds true for the first and second task in the test phase (Appendix \ref{appendix:study3_first_second}). While using AI in the exposure phase led to a higher rate of AI use on subsequent tasks compared to the control condition, doing tasks independently in the exposure phase did not lower the rate of AI use compared to the control condition: neither participants from condition 3 (completing 2 difficult task variants independently in the exposure phase) or condition 4 (completing 2 easy task variants independently in the exposure phase) used AI at a lower rate than participants from the control condition (condition 3 vs. control: $\beta=0.11$, $p=1$; condition 4 vs. control: $\beta=-0.10$, $p=1$).

AI exposure also exacerbated the speedup illusion. Although participants generally agreed that the easy task variants would be faster to complete independently, those who had previously completed tasks with AI reported lower agreement with the statement ``It was faster to complete the tasks independently than using AI''. We reverse coded the value to operationalize the speedup illusion ($\beta=-0.34$, $p<0.001$; Fig.~\ref{fig:study3_main}). We found no significant effect of previous completion mode on effort calibration ($\beta=0.08$, $p=0.28$). We confirm with qualitative analyses of people's justifications for their choice that time calibration played a paramount role in people's decision-making process: participants who completed tasks independently or used AI both reported that they did so because their choice was faster (Appendix \ref{appendix:study3_qualitative}).

Replicating our prior finding in Study 1 that AI assistance can backfire, we found that people who chose to use AI spent 7.06 seconds more than people who completed the tasks independently ($p<0.001$) and reported higher effort ($\beta=-0.11$, $p<0.001$)\footnote{After effect coding, independent completion $\text{AI use}=0$ is coded as $+0.5$, and AI-assisted completion is coded as $-0.5$; therefore, a negative coefficient means higher NASA-TLX were associated with AI-assisted completions. }. This finding is again robust to individual traits such as Need for Cognition (Appendix \ref{appendix:study3_stats}). Together, these findings show that prior AI use distort people's judgments about efficiency gains and carry over to subsequent choices, making people more likely to use AI.

\section{Discussion}\label{sec12}

We have shown that AI assistance may not actually provide the time and effort savings on simple tasks that people expect. We identify two calibration errors: a self-estimate miscalibration where people use AI at a rate significantly higher than the population-level predicted average and efficiency-gain illusions where people overestimate how much time and effort savings AI use can provide. We further show that AI exposure on just two tasks exacerbates miscalibration and leads to an increased rate of AI adoption on subsequent tasks. Our findings point to risks of a feedback loop of reliance where people habituate to using AI even when doing so does not save time or effort.


Our finding of an AI-specific speedup illusion builds on prior work identifying metacognitive failures in human-AI interaction, such as illusion of understanding and attributional ambiguity \citep{messeri2024artificial, kim2026llm}. Although we do not directly measure an individual's metacognitive monitoring of their AI use, we found a population-level miscalibration where people are generally unaware of their high rate of AI use on simple tasks. Moreover, while previous work typically measure only time or subjective mental effort in isolation \citep{lee2025impact,tamkinmccrory2025productivity},  we measure both to understand the distinctions between them. We uncover two distinct miscalibration patterns for time and effort: for time, people misjudge the efficiency of AI assistance, while for effort, they misevaluate how hard it would be to complete tasks independently. Studying both constructs together enabled us to uncover that AI use amplifies time miscalibration in particular, pointing to the possibility that time calibration is a main driving factor of whether to use AI.

The difference in the miscalibration patterns can be explained by distinct cognitive mechanisms behind time and effort forecasts. Time perception is shaped by experience pleasantness and cognitive load \citep{fraisse1984perception, fredrickson1993duration, zauberman2009discounting,liu1994mental}. AI is an emerging technology, and thus people may lack vivid representations of completing tasks with AI and therefore predict shorter durations \citep{matthews2016temporal}. On the other hand, for effort, prior work demonstrates a hedonic adaptation effect where people often overestimate the effort of independent work, and unaided tasks feel harder in anticipation \citep{brickman1971hedonic, brickman1978lottery, frederick199916}. Despite the different patterns, both miscalibrations lead people to overestimate the efficiency gain from AI assistance and potentially contributes to the high rate of AI use: people will use AI if they believe it to be faster, and people will avoid doing a task independently if they believe it to be strenuous. Our findings reveal different cognitive biases people may experience when making decisions about whether and how to engage with AI that provide insights about \textit{when} and \textit{why} people opt for AI use versus completing a task independently. 

We found that using AI does not seem to correct people's miscalibrations, but instead exacerbates them, directly increasing the likelihood of future AI reliance. In this cycle, these cognitive biases directly drive behaviors that entrench those same biases. The finding is particularly alarming because an excessive amount of offloading to AI has been linked to cognitive deskilling \citep{oktar2025identifying, gerlich2025ai, ahn2025preserving,hofman2023steroids, jose2025outsourcing, collins2025revisiting,shaw2026thinking}, overreliance, and disempowerment~\citep{ibrahim2025measuring,sturgeon2025humanagencybench,liu2026ai}. The practical consequence is compounding --- people use AI more than the predicted average rate, while simultaneously failing to recognize that doing so offers little efficiency benefit in terms of time and effort. Yet without awareness of this miscalibration, there is no incentive to change behavior \citep{elizondo2024self}. Our findings challenge prevailing narratives about AI-driven efficiency gains and underscore the urgency of examining how skewed mental models shape downstream behavior \citep{messeri2024artificial, bansal2019beyond, kelly2023capturing}.

Thus, addressing miscalibration is a promising intervention point to adjust behaviors and possibly mitigate overreliance \citep{collins2024modulating}. Our finding that independent completion of tasks did not lead to an increased rate of AI use compared to the control condition suggests that interventions at the belief level, instead of mere exposure or surface-level nudges, might be needed to encourage people to think independently to promote healthier and more efficient AI use. If AI is to truly empower people and even serve as a ``thought partner'' \citep{collins2024building}, it should not distort people's judgments about efficiency gains and foster overreliance; instead, people should be better calibrated so that they can make better-informed decisions about their AI use.

\section{Methods}\label{sec:methods}

\subsection{Problem Formulation}
While cognitive offloading is typically measured with time \citep{tamkinmccrory2025productivity, wang2025ai}, it is also related to the \textit{experience} of mental effort, which is fundamentally subjective \citep{steele2020perception}. Since time taken may not capture the subjective experience of mental effort, we measure both completion time as well as people's mental effort using the NASA-TLX index.

\subsection{Task Construction}
We constructed a total of 24 tasks (see Table \ref{tab:tasks}) based on a taxonomy of LLM use capturing four broad categories of cognitive skills required by different tasks: C1 -- Information Seeking, C2 -- Information Processing \& Synthesis, C3 -- Procedural guidance \& execution, and C4 -- Content creation \& transformation \citep{shelby2025taxonomy}.

\subsection{Experiment setup}

\subsubsection{Study 1}

For Study 1 ($N=498$), each participant completed a total of four tasks, with the option to prompt an AI chatbot (GPT-4o) for assistance on each task; the chatbot was embedded in the survey interface below the box for participants to type their answer. We consider a participant to use AI on a task if they enter at least one prompt.

For each task, task difficulty (easy vs. hard) was randomized at the task level and was tested as a moderator (Appendix \ref{appendix:study1_stats}). We recorded the completion time with a hidden timer and asked participants to report the subjective mental effort using the NASA-TLX index after completing each task (full details in Appendix \ref{appendix:task_details}). At the end of the survey, participants completed the Need For Cognition scale (NFC) and the AI Assessment scale (AIAS) \citep{grassini2023development}. We analyze their effects in Appendix \ref{appendix:study1_individual}.

\textbf{Participants.} Participants are recruited through the Prolific platform, comprising a representative sample of the US adult population. Our sample size is N=498 in total.\footnote{We conducted power analyses based on effect size estimates from pilot data ($\alpha=0.05$, $\text{power}=0.80$) and determined that $N=498$ provides adequate power to detect between-subject differences between the predicted vs. actual rates of AI use} The participants were 51\% female and 49\% male; 64\% white, 12\% Black, 11\% mixed, and 6\% Asian.

\subsubsection{Study 2} \label{methods:study2}

To measure people's completion times and how their expectations for them differed from reality, we collected two separate samples: a prediction sample and a completion sample. The samples are collected separately on Prolific, with 307 overlapping participants. 

\paragraph{Prediction sample}  In the prediction sample ($N=600$), participants were randomly assigned to one of two hypothetical conditions, making predictions for their own independent completions and for assisted (AI or another participant) completions. To avoid biases from prediction order, we further randomized participants to either first predict effort then time or first time then effort (more details in Appendix \ref{appendix:prediction_order}). To determine whether the difference observed above is unique to AI assistance, we used ``another participant'' (participants are told that this other participant is highly intelligent) as a baseline. Without having to complete the task while having a detailed task description, a participant was first asked to state whether they would like to complete the task independently or with the external assistance of AI/human. Then, participants were asked to predict how much time and effort (in the assigned order) it would take to complete the task independently and with AI assistance or with assistance from another participant. After making predictions, participants completed NFC and AIAS measures as in Study 1.

\paragraph{Completion sample} In the completion sample ($N=1001$), participants were randomly assigned to complete tasks either independently or with AI assistance. For the AI assistance condition, we provided an embedded chat interface with GPT-4o on the task page that participants needed to prompt at least once before being able to proceed to the next page. In the independent completion condition, participants were not allowed to use any form of AI, and copying and pasting was disabled. Participants were presented a randomized mix of easy and difficult task variants. Each task was completed by around 84 participants for each condition. We used a hidden timer to record the completion time for each task. After the completion of each task, participants answered the NASA-TLX. To understand people's calibration and efficiency gain when it does not affect the outcome, we annotated all final answers and filtered out ones that were incorrect and low-quality. In general, participants made a decent effort to complete the tasks across both conditions; we excluded 4.74\% of the responses in the independent condition and 2.32\% in the AI assistance condition.

\textbf{Participants.} Participants are recruited through the Prolific platform, comprising a representative sample of the US adult population. Our sample size is N=1601 in total: 600 in the prediction sample and 1001 in the completion sample\footnote{We conducted power analyses using effect size estimates from pilot data ($\alpha = 0.05$, $\text{power}=0.80$) for predicted independent vs. predicted AI, predicted AI vs. actual AI (time and effort), and the tests for the efficiency-gain illusions, and determined that $N=600$ for the prediction sample and $N=1000$ for the completion sample would provide adequate power for all main analyses.}. The participants were 51\% female, 49\% male; 64\% white, 12\% Black, 11\% mixed, and 6\% Asian. 

\subsubsection{Study 3}

Our third preregistered study (N=592) investigates how previous exposure to different task completion modes affects people's decisions to use AI to complete simple cognitive tasks. Study 3 consists of two phases: an exposure phase and a test phase. In the exposure phase, we randomly assigned participants to one of five possible conditions: condition 1 --- completing two difficult task variants using AI, condition 2 --- completing two easy task variants using AI, condition 3 --- completing two difficult task variants independently, condition 4 --- completing two easy task variants independently, or condition 5 --- not completing any tasks and instead just answering the full version of the Big Five (control condition). In the exposure phase, participants must complete the first two tasks (or no tasks) using their assigned method. The easy task variants were randomized from a pool of 6 easy tasks used in Study 1 and Study 2, and the difficult task variants are randomized from a pool of the difficult version of these tasks (see Table \ref{tab:tasks} for the tasks used).

In the test phase, participants completed two easy task variants with the option to use AI for help, similar to the interface of Study 1. The tasks again came from the pool of 6 possible tasks. All tasks were randomized such that each task (its easy or difficult variant) only appeared once. We again recorded completion times and the self-reported NASA-TLX. After completing the last task, we asked participants to self-report the perceived difficulty, their time calibration (i.e. by answering how much they agree with the statement ``It is faster to complete the task independently than using AI''), their effort calibration (i.e. by answering how much they agree with the statement ``It would require more mental effort to complete the task independently than using AI'') and confidence on a 7-point scale. Then, we asked participants to explain why they chose to complete the task independently or using AI and coded their responses qualitatively (see Appendix \ref{appendix:study3} for more details). Similar to Study 1 and 2, we asked participants to complete the NFC and AIAS scales at the end of the survey.

\textbf{Participants.} Participants are recruited through the Prolific platform, comprising a representative sample of the US adult population. Our sample size is $N=592$ in total\footnote{We conducted power analyses testing how previous completion modes (AI vs. independent) predicts subsequent AI adoption rate using effect size estimates from pilot data ($\alpha = 0.05$, $\text{power}=0.80$) and determined that $N=490$ provides adequate power.}. The participants were 51\% female, 49\% male; 64\% white, 12\% Black, 11\% mixed, and 6\% Asian.

\subsection{Limitations and Future Directions} 

Our study has various limitations that point to opportunities for future work. First, we focus on tasks that can be completed in under 5 minutes; an interesting direction of future work is to identify the specific complexity level at which AI assistance genuinely saves time and effort. Moreover, AI capabilities are changing over time, which affects people's decision to defer a task, the amount of oversight needed, and how quickly AI can complete each task. Nevertheless, our study focuses on human miscalibration rather than AI capability. Future work should explore how human calibration changes over time as capabilities change and as people spend more time working with AI. 

Second, our experiment did not directly control for participant motivation or incentives. Although we manually went through the responses and excluded low-effort and incorrect responses, crowdworking contexts may impact  task completion time, response quality, and AI adoption rates compared to other settings with more personal stakes \citep{yurt2024questionnaire}. Nonetheless, this setting enables us to understand overall miscalibrations in AI use. Future work should examine how incentive structures and motivation levels (e.g. answer quality) interact with the completion times and reported effort. 

Third, there are individual variations in AI use (Appendix \ref{appendix:study2_behavior}). Participants used AI in very different ways, and the degree of cognitive effort required even within the AI condition could vary. For example, it is possible that participants prompted the model but came up with a different final answer, regardless of what the AI response was. Some participants only used AI to double check their answer or prompted AI out of curiosity without using the responses. Therefore, the binary variable of AI use in Study 1 and Study 3 is a rough proxies for the choice of using AI and does not capture \textit{how} participants used AI. We provide a qualitative analysis of how participants used AI in Appendix \ref{appendix:study2_behavior}.

Fourth, in Study 3, we measured AI adoption immediately after the AI exposure phase. This limits conclusions about longer-term behavioral change. Future work should include longitudinal analyses to directly examine how AI use affects longer-term behavioral outcomes and AI use patterns \citep{collins2024modulating}. Our study 3 finding could be due to persistence and the inertia effect, where participants use AI because they had been previously exposed to doing so \citep{alos2016inertia}. Regardless of the limitation, our results show how the inertia could play out in real life and have the same effect on people. Future work should study how it carries through in long-term settings to capture how beliefs about calibrations shape AI-use behaviors, and how behaviors change subsequent beliefs longitudinally. 

Finally, Study 1 and Study 2 rely on between-subjects aggregates to avoid carryover biases, and we do not directly compare the same individual's stated vs. actual rate of AI use or their predicted vs. actual completion times and effort. The prediction sample and completion sample were also not completely disjoint, with some participants completing both samples on Prolific. Future work should examine within-subject patterns to directly study metacognitive monitoring around AI use while minimizing biases in the dependent variables.

\backmatter

\bmhead{Acknowledgements}

We would like to thank members of the Social Interaction Lab (SoIL) and the Jurafsky lab for their valuable feedback on the project. We thank Polina Tsvilodub and Charley Wu for their helpful feedback on an earlier version of the draft. Additionally, we thank Lujain Ibrahim and Vishakh Padmakumar for helpful discussions and Katherine Lu for visualization feedback.

\bmhead{Ethics approval and consent to participate}
Our study was approved by the Stanford Institutional Review Board (IRB) under protocol 83204. The reviewing committe determined that the only involvement of human subjects in the research activities will be in one or more of the categories that are exempt from the regulations at 45 CFR 46 or 21 CFR 56. Informed consent was obtained from all human research participants. 

\bmhead{Code and data availability} 
Data and code from the study are both available at \href{https://osf.io/v6zgy}{this link}.

\newpage
\FloatBarrier
\begin{appendices}

\section{Task details}\label{appendix:task_details}

Our study was approved by the Stanford Institutional Review Board (IRB) under protocol 83204 and
pre-registered at \href{https://osf.io/xp3td}{Study 1}, \href{https://osf.io/8x9j6}{Study 2}, and \href{https://osf.io/7ygmw}{Study 3}. For the full list of tasks used, see Table \ref{tab:tasks}.

\begin{table}[htbp]

\begin{tabular}{
p{2.5cm}   
p{0.8cm} 
p{5cm} 
p{5cm} 
}
\toprule
\textbf{Category} & \textbf{Q\#} & \textbf{Easy Task (G1)} & \textbf{Difficult Task (G2)} \\
\midrule

\multirow{3}{3cm}{Information Seeking}
& 1 & Name the color that you get when you mix white and black. 
& Name the color that you get when you mix red, orange, yellow, green, blue, and purple. \\

& 2 & \textbf{Name one Olympic medalist. }
& \textbf{Name five Olympic medalists.} \\

& 3 & \textbf{Name another word for “enjoy”.} 
& \textbf{Name five other words for “enjoy”.} \\

\midrule

\multirow{3}{3cm}{Information Processing}
& 1 & Provide instructions on how to boil an egg in 2 sentences. 
& Provide detailed instructions on how to make a salad that includes a hard boiled egg. \\

& 2 & Read through the following article and summarize it in one sentence: ``On the last night before departure, Mara packed memories instead of clothes...'' (50 words)  
& Read through the following article and summarize it in one sentence: ``The town of Alder Creek had a clock that never told the same time twice...'' (200 words) \\

& 3 & \textbf{Would you pull a lever to divert a runaway trolley, saving five people and not harming anyone? Write a brief justification in three sentences.  } 
& \textbf{There’s a trolley traveling down a path upon which lie ten people. These ten people are promising young minds, all studying medicine with the hopes of serving their town. All of a sudden, you’re given control of this trolley... }  \\

\midrule

\multirow{3}{3cm}{Procedural Guidance and Execution}
& 1 & \textbf{Find the spelling error in the following sentence: ``The quick borwn fox jumps over the lazy dog.''}
& \textbf{Find the spelling error in the following text: ``... In the kitchen, the aroma of freshly brewed cofee filled the air, mingling with the faint scent of rain from the night before...'' (134 words).} \\

& 2 & \textbf{If today is Monday, what day in 2 days?} 
& \textbf{If this month is April, what month in 15 months?} \\

& 3 & \textbf{Calculate: $27+56$.} 
& \textbf{Calculate: $27+56+32+76+189+672+328$.} \\

\midrule

\multirow{3}{3cm}{Content Creation}
& 1 & Write a short thank-you message for someone who went to your birthday party.
& Write a short thank-you for someone who went to yoru birthday party. Be specific about what you appreciate about their presence and how much you like their present. \\

& 2 & Write a one-sentence product review of a water bottle. 
& Write a 5-sentence product review of a water bottle, covering pros, cons, and how it compares to alternatives.  \\

& 3 & Rewrite this text to be clearer and more readable: ``When I have nothing to do, I like to do nothing but read anything at all.'' 
& Rewrite this text to be clearer and more readable: ``When I have nothing to do, I like to do nothing but read anything at all. This trait makes me fun to people cause they like me. I also like to eat lots of food. Lots. I like food that makes me fat. Sugary food! But it's something that also makes people love me. I am lovable. I am a lovable person.'' \\

\bottomrule
\end{tabular}

\caption{\textbf{All tasks are used in Study 1 and Study 2.} The subset of tasks used for study 3 are highlighted in \textbf{bold}. We denote the four categories as C1, C2, C3, and C4 and each question under the category as Q1, Q2, and Q3. Easy tasks are denoted as G1, and difficult tasks are denoted as G2. There are 24 tasks in total, 2 per each of the 12 task types.  }
\label{tab:tasks}
\end{table}

\begin{figure}[t]
    \centering
    
    \begin{subfigure}[t]{0.48\columnwidth}
        \centering
        \includegraphics[width=\linewidth]{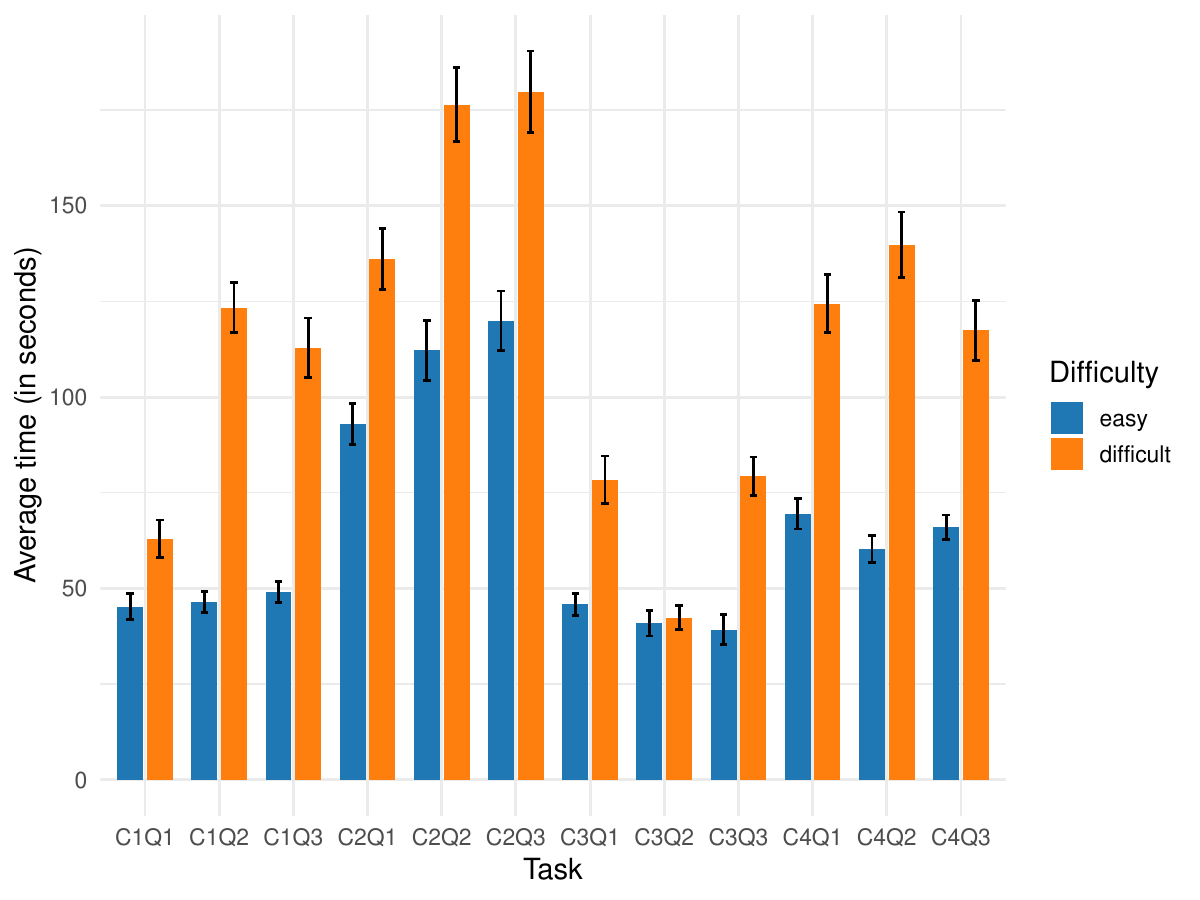}
    \end{subfigure}
    \hfill
    \begin{subfigure}[t]{0.48\columnwidth}
        \centering
        \includegraphics[width=\linewidth]{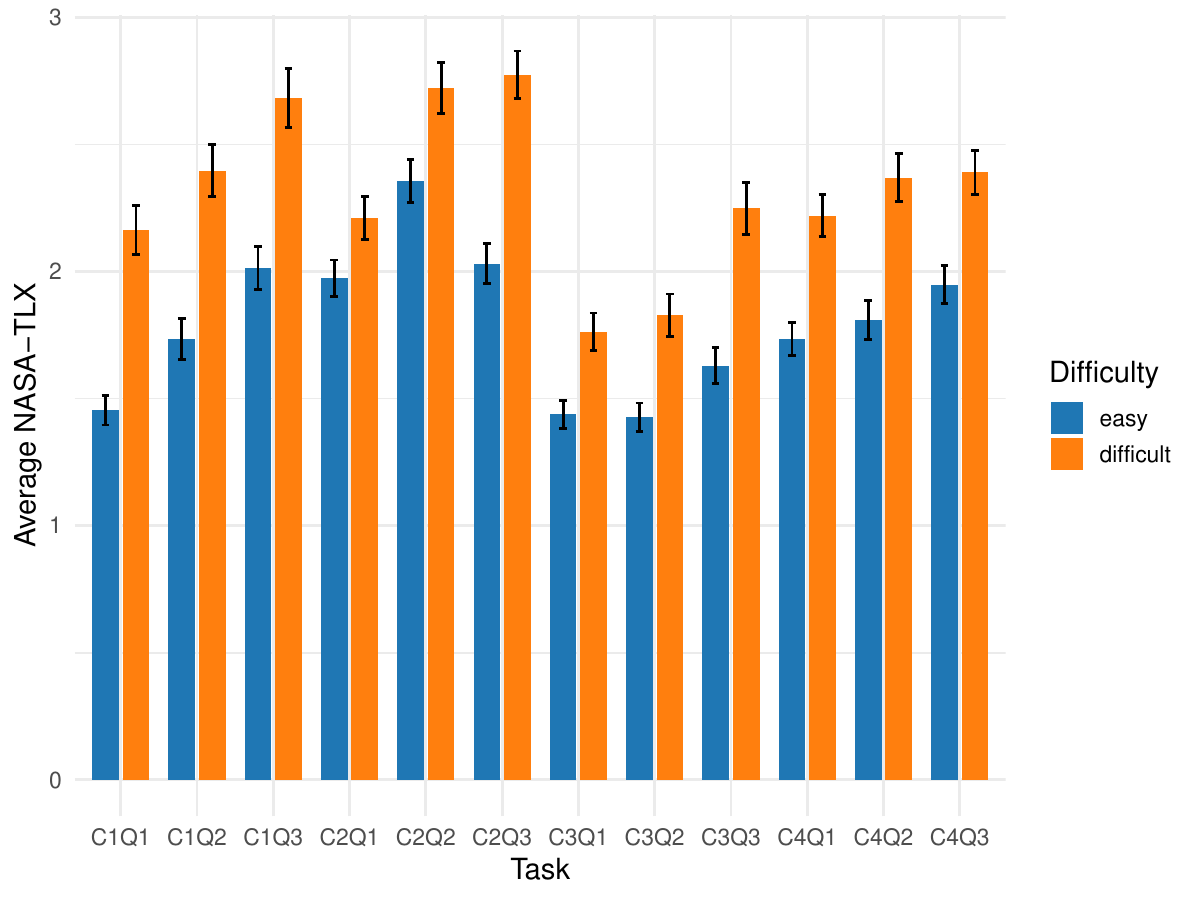}
    \end{subfigure}
    
    \caption{(a). \textbf{Completion times} (independent and AI condition combined) by tasks after filtering out low-quality and incorrect responses. We confirm that for all but one task, participants spent more time completing the difficult version of a task than the easy one. (b). \textbf{NASA-TLX (average) by tasks} (independent and AI condition combined), after filtering out low-quality and incorrect responses. We confirm that for all tasks, people experienced more cognitive effort completing the difficult version than the easy version.}
    \label{fig:appendix_time_effort}
\end{figure}

\subsection{Measurement details} \label{appendix:nasa}

\paragraph{NASA-TLX} 
Table \ref{tab:appendix_nasa_tlx} shows the full NASA-TLX scale. We report the average score across all 5 items, with item 3 reverse coded. Each item was on a 7-point scale, with 1 labeled as ``Very Low'', and 7 labeled as ``Very High''.

\begin{table}[h]
\centering
\begin{tabular}{lll}
\hline
\textbf{Item \#} & \textbf{Abbrev} & \textbf{Question} \\
\hline
Item 1 & mental demand & How mentally demanding was completing the task for you? \\
Item 2 & hurried & How hurried or rushed was the pace of the task for you? \\
Item 3 & successful & How successful were you in accomplishing what you were asked to do? \\
Item 4 & hard work & How hard did you have to work to achieve your level of performance? \\
Item 5 & insecure & How insecure, discouraged, irritated, stressed, and annoyed were you? \\
\hline
\end{tabular}
\caption{NASA-TLX Items used in the study. There are 5 items in total; we will use the abbreviation in future tables to represent each corresponding item.}
\label{tab:appendix_nasa_tlx}
\end{table}

\paragraph{Need For Cognition Scale}
Table \ref{tab:appendix_nfc} displays the Need For Cognition scale \citep{cacioppo1982need} for measuring participants' willingness to think. Each question was presented on a 5-point scale, with 1 labeled as ``Extremely Uncharacteristic'' and 5 labeled as ``Extremely characteristic''. Participants are asked to select the number that indicates the extent to which they feel is characteristic of them. We averaged the items where item 3 and 4 are reverse coded.

\begin{table}[h]
\centering
\begin{tabular}{lp{10cm}}
\hline
\textbf{Item \#} & \textbf{Statement} \\
\hline
Item 1 & I would prefer complex to simple problems. \\
Item 2 & I like to have the responsibility of handling a situation that requires a lot of thinking. \\
Item 3 & Thinking is not my idea of fun. \\
Item 4 & I would rather do something that requires little thought than something that is sure to challenge my thinking abilities. \\
Item 5 & I really enjoy a task that involves coming up with new solutions to problems. \\
Item 6 & I would prefer a task that is intellectual, difficult, and important to one that is somewhat important but does not require much thought. \\
\hline
\end{tabular}
\caption{Individual items for the Need For Cognition scale.}
\label{tab:appendix_nfc}
\end{table}

\paragraph{AI Assessment Scale (AIAS)}
The AIAS scale measures participants' attitudes towards AI. Each question was presented on a 10-point scale, with 1 labeled as ``Not At All'' and 10 labeled as ``Completely Agree''. There are four items in the scale: Item 1 ``I believe that AI will improve my life,'' Item 2 ``I believe that AI will improve my work,'' Item 3 ``I think I will use AI technology in the future,'' and Item 4 ``I think AI technology is positive for humanity.'' The correlation between NFC and AIAS is minimal across the data in our four samples: 0.08 in Study 1 prediction sample; 0.07 in Study 1 completion sample; 0.19 in Study 2 sample; and 0.05 in Study 3 sample.

\section{Study 1 Details} \label{appendix:study1}

\subsection{Statistical Testing} \label{appendix:study1_stats} We fit the model $\text{AI use} \sim \text{type} * \text{difficulty} + \text{NFC} + (1  \mid \text{participantID}) + (1 \mid \text{taskID})$ to compare the predicted rate of AI use with the actual rate of AI use. In this model, type is either predicted (data from Study 2 prediction sample) or actual (people's actual rate of AI use from Study 1). AI use is coded as 1 in the prediction sample if participants chose AI for the stated preference question (``If you need to complete the task, how would you prefer to complete it? A: Independently B: With the help of an AI model''), and coded as 0 if participants chose ``independent''.  It is is coded as 1 in the Study 1 sample if participants used AI at least once for the task, and 0 if not.

\subsection{Variations Across Categories and Tasks}\label{appendix:study1_categories}

\begin{figure}[t] 
    \centering
    \includegraphics[width=\columnwidth]{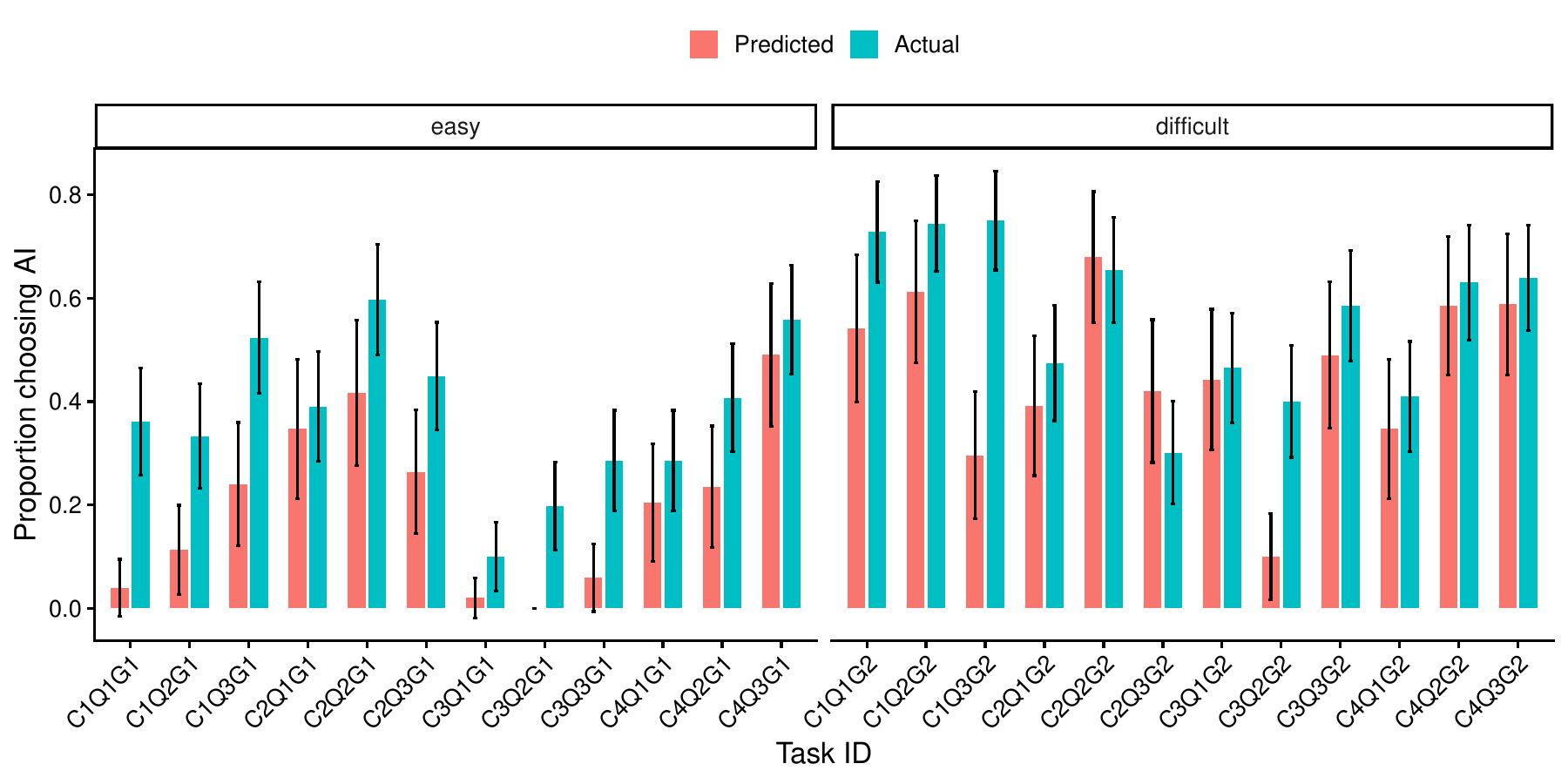} 
    \caption{\textbf{For most easy task variants, the actual rate of AI use is significantly greater than the predicted one.} The predicted vs. actual rates of AI use on each of the 12 tasks within each difficulty level. The task IDs denote the category (first two digits, e.g. C1), question ID (second two digits, e.g. Q1), and difficulty level (last two digits, e.g. G1). For many easy task variants, the predicted rate is close to 0 while the actual rate is significantly greater.}
    \label{fig:study2_task}
\end{figure}

For all four categories, the actual rate of AI use is significantly greater than the stated rate of AI use: C1 (Information Seeking; $\beta=2.01$, $p<0.001$), C2 (Information Processing; $\beta=0.51$, $p<0.05$), C3 (Procedural Guidance and Execution; $\beta=1.53$, $p<0.01$), and C4 (Content Creation; $\beta=0.63$, $p<0.01$). For each task, we compare the predicted rate of how a participant would like to complete the task with the actual rate: the difference is significant for 8 out of 12 easy task variants, and 3 out of 12 difficult task variants, again highlighting the susceptibility of easy task variants to the self-estimate miscalibration.

\begin{table}[ht]
\centering
\caption{Individual tasks with a significant difference between stated and actual rates of AI use}
\label{tab:appendix_study1_stats}
\begin{tabular}{lcccc}
\hline
\textbf{Task ID} & \textbf{Estimate} & \textbf{SE} & \textbf{z-ratio} & \textbf{p-value} \\
\hline
\multicolumn{5}{c}{\textit{Easy Tasks}} \\
\hline
C1Q1G1 & 0.35 & 0.07 & 4.60 & $<.0001$ \\
C1Q2G1 & 0.26 & 0.07 & 3.46 & $0.0005$ \\
C1Q3G1 & 0.32 & 0.08 & 4.16 & $<.0001$ \\
C2Q2G1 & 0.16 & 0.08 & 2.08 & $0.04$ \\
C2Q3G1 & 0.18 & 0.07 & 2.50 & $0.01$ \\
C3Q2G1 & 0.25 & 0.07 & 3.30 & $0.001$ \\
C3Q3G1 & 0.23 & 0.08 & 3.09 & $0.002$ \\
C4Q2G1 & 0.19 & 0.07 & 2.54 & $0.01$ \\
\hline
\multicolumn{5}{c}{\textit{Difficult Tasks}} \\
\hline
C1Q2G2 & 0.19 & 0.08 & 2.54 & $0.0112$ \\
C1Q3G2 & 0.47 & 0.07 & 6.28 & $<.0001$ \\
C3Q2G2 & 0.33 & 0.08 & 4.36 & $<.0001$ \\
\hline
\end{tabular}
\end{table}

Table \ref{tab:appendix_study1_stats} shows the tasks with a significant difference between the stated and actual rates of AI use. We highlight a few examples below: for the task on coming up with another word for ``enjoy'' (C1Q3G1), the actual rate of AI use is as high as 54\% ($\beta=0.32$, $p<0.001$) even though almost no one stated they would use AI for this task. For the question ``If today is Monday, what day will it be in two days?'' (C3Q2G1), again almost no participants stated that they would use AI to do it, but more than 21\% of participants actually used AI ($\beta=0.25$, $p<0.01$);

\subsection{Individual Differences} \label{appendix:study1_individual}
It is possible that AI use is largely driven by individual traits and preferences, such as one's tendency to enjoy thinking. We use the Need for Cognition scale to measure a participant's willingness to think and hypothesize that higher NFC values (i.e. more likely to enjoy thinking) predicts a lower rate of AI use. To investigate whether Need For Cognition predicts people's likelihood of using AI on a task, we fitted the model $\text{AI use} \sim   \text{NFC}*\text{difficulty} + (1  \mid \text{participantID}) + (1 \mid \text{taskID})$ and did not find any significant effects of NFC ($\beta=-0.07$, $p=0.50$). 

However, we found that the AIAS average score predicts AI use: the higher the value is (i.e. more optimistic about AI), the more likely participants were to use AI on a task ($\beta=0.26$, $p<0.001$). We found that all four items on the scale positively predicts AI use (item 1: $\beta=0.24$, $p<0.001$; item 2: $\beta=0.23$, $p<0.001$; item 3: $\beta=0.19$, $p<0.001$; item 4: $\beta=0.21$, $p<0.001$). The results show that general attitudes towards AI can better predict people's likelihood of using AI on tasks than people's propensity to think.

\subsection{Excluding Extreme Data} \label{appendix:study1_behavior}

It is possible that some participants always used AI or never used AI because of personal preference. We found that out of all participants, 17.5\% used AI for all four tasks, while 20.7\% participants never used AI.

\begin{figure}[t] 
    \centering
    \includegraphics[width=\columnwidth]{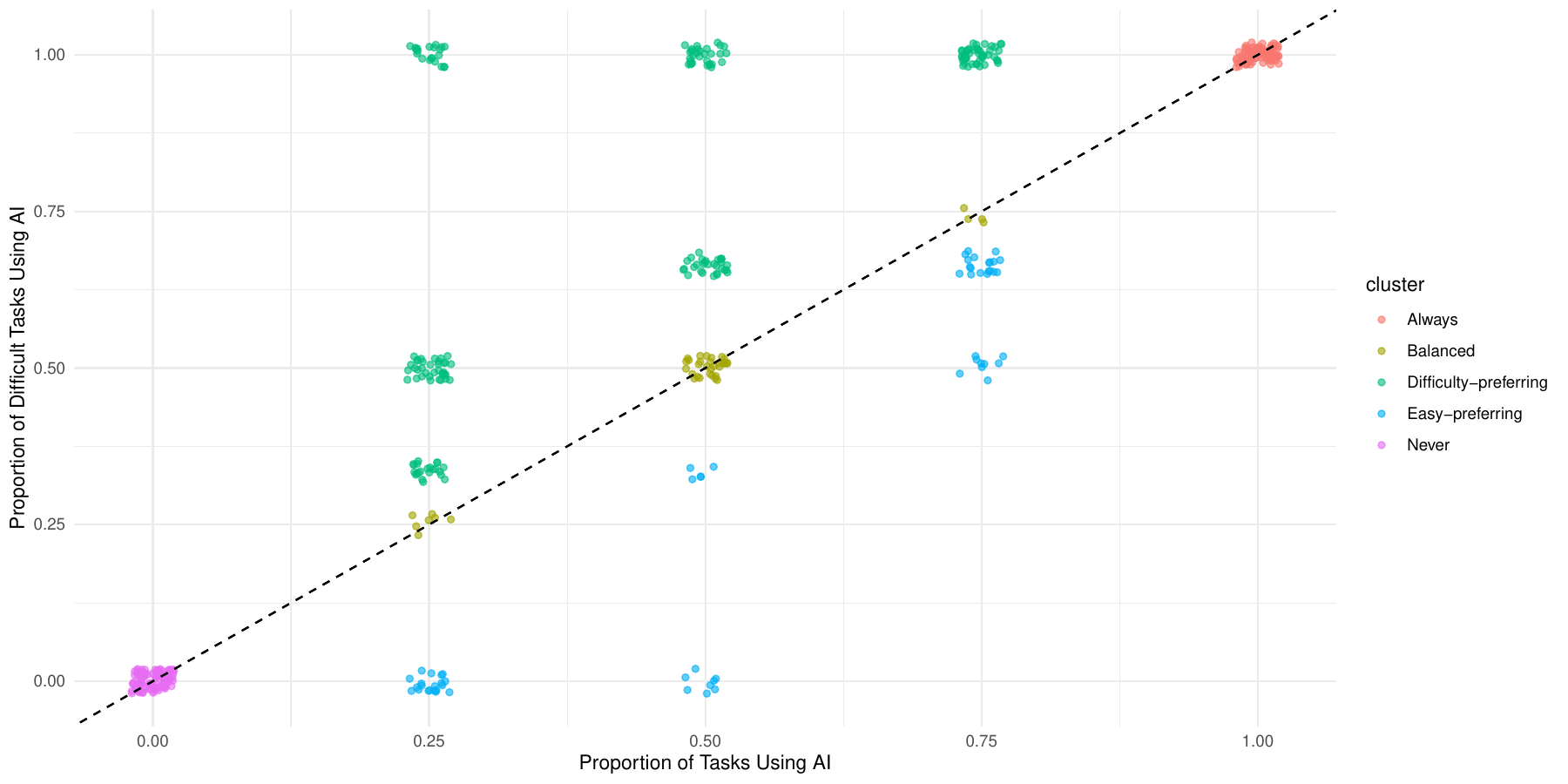} 
    \caption{\textbf{Distinct clusters of AI-use patterns emerge}. The x-axis shows the proportion of tasks each participant used AI for, and y-axis is the proportion of difficult task variants the participants used AI for. More participants chose to use AI on difficult task variants as there are more points above the dotted diagonal line. The pink cluster on the bottom-left shows participants who never uses AI, and the red cluster on the top-right shows participants who uses AI on all the tasks.}
    \label{fig:study2_AI_use}
\end{figure}

We exclude these two types of outlier data and focus on the selective use of AI --- when participants used AI once, twice, or three times out of the all four tasks. We filtered from 1992 task completions to 1232 and performed the same set of analyses, comparing the actual behavioral data with the stated rate of preferences. We replicated the finding that the likelihood of using AI is significantly greater than what people predicted across all tasks ($\beta=0.78$, $p<0.001$), even after removing the outlier data. The differences are again significant for both easy ($\beta=0.90$, $p<0.001$) and difficult task variants ($\beta=0.67$, $p<0.001$). However, the interaction with difficulty level is no longer significant ($\beta=0.23$, $p=0.22$). One possible explanation is that the mindless use of AI on easy task variants of participants who always uses AI contributed to the higher rate of difference between easy and difficult task variants. Figure \ref{fig:study2_AI_use} provides a visualization of the different behavioral patterns: we identify a cluster of participants who never used AI (pink), a cluster of participants who always used AI (red), and more participants who selectively used AI for difficult task variants than participants who use AI for easy task variants (green).

\section{Study 2 details} \label{appendix:study2}

\subsection{Statistical Tests} \label{appendix:study2_stats} 
In this section, we provide details on the models fitted to test each hypothesis for Study 2. To compare how the predicted time differed across conditions, we fitted the following model: $\text{time} \sim \text{prediction target} * \text{group} + \text{order} + (1 + \text{prediction target}\mid \text{participantID}) + (1 \mid \text{taskID})$. Prediction target is either independent or assistance, and group refers to the source of assistance (AI or another participant). Order refers to the order in which participants made predictions (either predicting effort before time, or predicting time before effort). We found that in this regression, the order of prediction has no effect on the predicted time ($\beta=-4.90$,  $p=0.15$). To compare the predicted time between independent and AI-assisted conditions, we examined the simple effect of prediction target when the group is AI; to compare the predicted time between independent and another-participant-assisted conditions, we examined the simple effect of prediction target when group is another participant. To compare how the predicted effort differs across conditions, we replaced the outcome variable with effort (NASA-TLX average) and performed the same statistical tests.

To understand people's calibration, we combined the prediction data with the completion data, adding another variable ``type'' to specify the data source (prediction vs. completion). Then, we fitted the following model to predict time, where time is either the predicted time or the actual completion time: $\text{time} \sim \text{condition} * \text{type} + \text{difficulty} + (1 \mid \text{participantID}) + (1 \mid \text{taskID})$. Condition refers to whether people predicted or completed for independent or AI-assisted conditions. We effect coded condition, type, and difficulty. To examine the calibration patterns for each condition, we looked at the simple effect of type at either the independent or AI-assisted conditions. Similarly, for effort calibration, we changed the outcome variable from time to NASA-TLX (average) and performed the same statistical tests. To determine whether people overestimate the time (or effort) reduction between the predicted and actual ones (corresponding to Equation~\eqref{equation:time_delta} and Equation~\eqref{equation:effort_delta}), we examine the interaction between condition and type in the two models, respectively. 

Finally, to compare the actual time between independent and AI-assisted conditions, we used data from the completion sample to fit the model below: $\text{time} \sim \text{condition} * \text{difficulty}  + (1 \mid \text{participantID}) + (1 \mid \text{taskID})$, where condition is independent or AI assistance. We again effect coded condition and difficulty. To understand whether there is a difference in completion time across conditions, we examine whether there is an effect at the condition variable. Similarly, to determine whether the mental effort differs across conditions, we again fited the same model and changed the outcome variable from time to NASA-TLX (average).

\subsection{Results across task types, difficulty levels, and individual tasks} \label{appendix:study2_variations}

We found that AI assistance significantly reduces mental effort for 17 out of 24 tasks. Overall, the NASA-TLX average (higher means more effortful) lowers from 2.35 to 1.76, by 0.59 points ($p<0.001$) across all tasks. 

To gain a more nuanced understanding of the variations and generalizability of the findings across different categories of tasks, difficulty levels, and specific tasks, we repeated the analyses reported above at a category, difficulty, and task level. To test calibration for time across categories, we changed the model to: \\

$\text{time} \sim \text{condition} * \text{type} + \text{type}*\text{category} + (1 \mid \text{participantID}) + (1 \mid \text{taskID})$, \\
where category is the four task categories. We examine the simple effects of type per category. To test calibration for time across difficulty levels, we changed the model to: \\

$\text{time} \sim \text{condition} * \text{type} + \text{type}*\text{difficulty} + (1 \mid \text{participantID}) + (1 \mid \text{taskID})$, \\
We examine the simple effects of type per difficulty level.

Even though overall, we did not find a statistically significant difference between the predicted versus actual completion times for the independent condition, there are some variations across the task categories. For C1 tasks ($\beta=-9.56$, $p<0.05$) and C4 ($\beta=-21.62$, $p<0.001$) tasks, participants' predicted times are greater than the actual completion times. On easy task variants, participants also predicted their completion times to be significantly greater than the actual completion time ($\beta=-8.93$, $p<0.05$). The predicted time is significantly greater than the actual time for the AI assistance condition across all four task categories and both difficulty levels, with the statistical testing results shown in Table \ref{tab:appendix_categories_AI}. For effort calibration, we found that the overall findings hold across all four task categories and both difficulty levels -- no significant difference between predicted and actual NASA-TLX for the AI condition, and an overestimation of actual effort for the independent condition (see Table \ref{tab:appendix_effort}). 

\begin{table}[h]
\centering
\begin{tabular}{lcccc}
\hline
\textbf{Group} & \textbf{Estimate} & \textbf{SE} & \textbf{z-ratio} & \textbf{p-value} \\
\hline
C1 & 38.9 & 4.30 & 9.05 & $<.0001$ \\
C2 & 65.1 & 4.30 & 15.14 & $<.0001$ \\
C3 & 41.0 & 4.31 & 9.50 & $<.0001$ \\
C4 & 26.8 & 4.29 & 6.27 & $<.0001$ \\
Easy & 39.4 & 3.66 & 10.75& $<.0001$ \\
Difficult & 46.6 & 3.67 & 12.67 & $<.0001$ \\
\hline
\end{tabular}
\caption{Difference between the actual vs. predicted time for the AI-assisted condition across task categories and difficulty levels.}
\label{tab:appendix_categories_AI}
\end{table}

\begin{table}[h]
\centering
\begin{tabular}{lcccccccc}
\hline
 & \multicolumn{4}{c}{\textbf{Independent}} & \multicolumn{4}{c}{\textbf{AI}} \\
\textbf{Group} & \textbf{Est} & \textbf{SE} & \textbf{z} & \textbf{p} 
               & \textbf{Est} & \textbf{SE} & \textbf{z} & \textbf{p} \\
\hline
C1 & -0.18 & 0.06 & -3.12 & 0.002 
   & 0.10 & 0.06 & 1.65 & 0.10 \\

C2 & -0.39 & 0.06 & -6.72 & $<.0001$ 
   & -0.11 & 0.06 & -1.85 & 0.04 \\

C3 & -0.29 & 0.06 & -4.96 & $<.0001$ 
   & -0.01 & 0.06 & -0.17 & 0.86 \\

C4 & -0.35 & 0.06 & -5.96 & $<.0001$ 
   & -0.07 & 0.06 & -1.10 & 0.27 \\

Easy & -0.29 & 0.05 & -5.79 & $<.0001$ 
     & -0.009 & 0.05 & -0.17 & 0.87 \\

Difficult & -0.32 & 0.05 & -6.30 & $<.0001$ 
          & -0.04 & 0.05 & -0.68 & 0.50 \\
\hline
\end{tabular}
\caption{Difference between the NASA-TLX average for completion and prediction (completion $-$ prediction) for both independent and AI-assisted conditions across task categories and difficulty levels. Negative values indicate underestimation of completion time relative to prediction.}
\label{tab:appendix_effort}
\end{table}

To test calibration for time across specific tasks, we used the following model: \\

$\text{time} \sim \text{condition} * \text{taskID}  + (1 \mid \text{participantID})$, \\
We examine the simple effects of type per category. 

\begin{figure}[t]
    \centering
    
    \begin{subfigure}[t]{0.48\columnwidth}
        \centering
        \includegraphics[width=\linewidth]{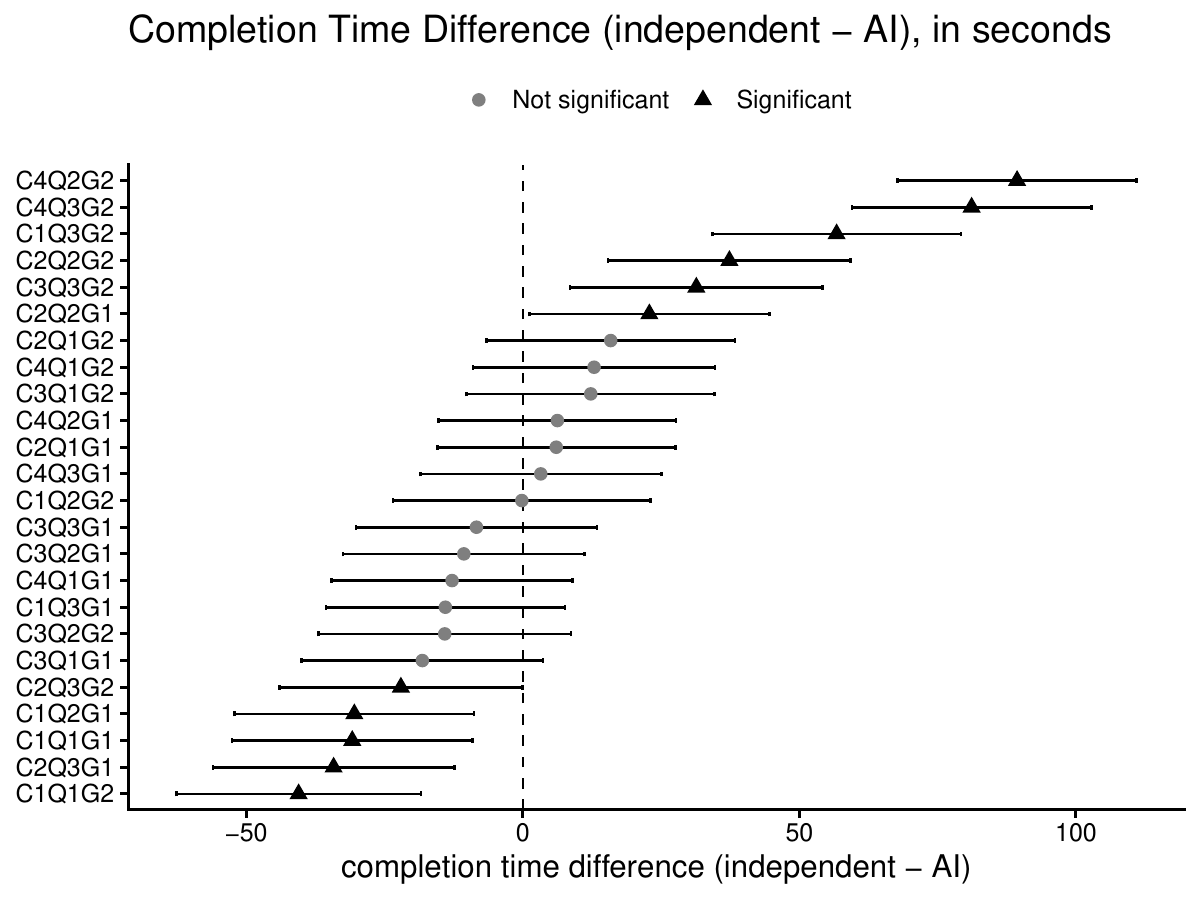}
    \end{subfigure}
    \hfill
    \begin{subfigure}[t]{0.48\columnwidth}
        \centering
        \includegraphics[width=\linewidth]{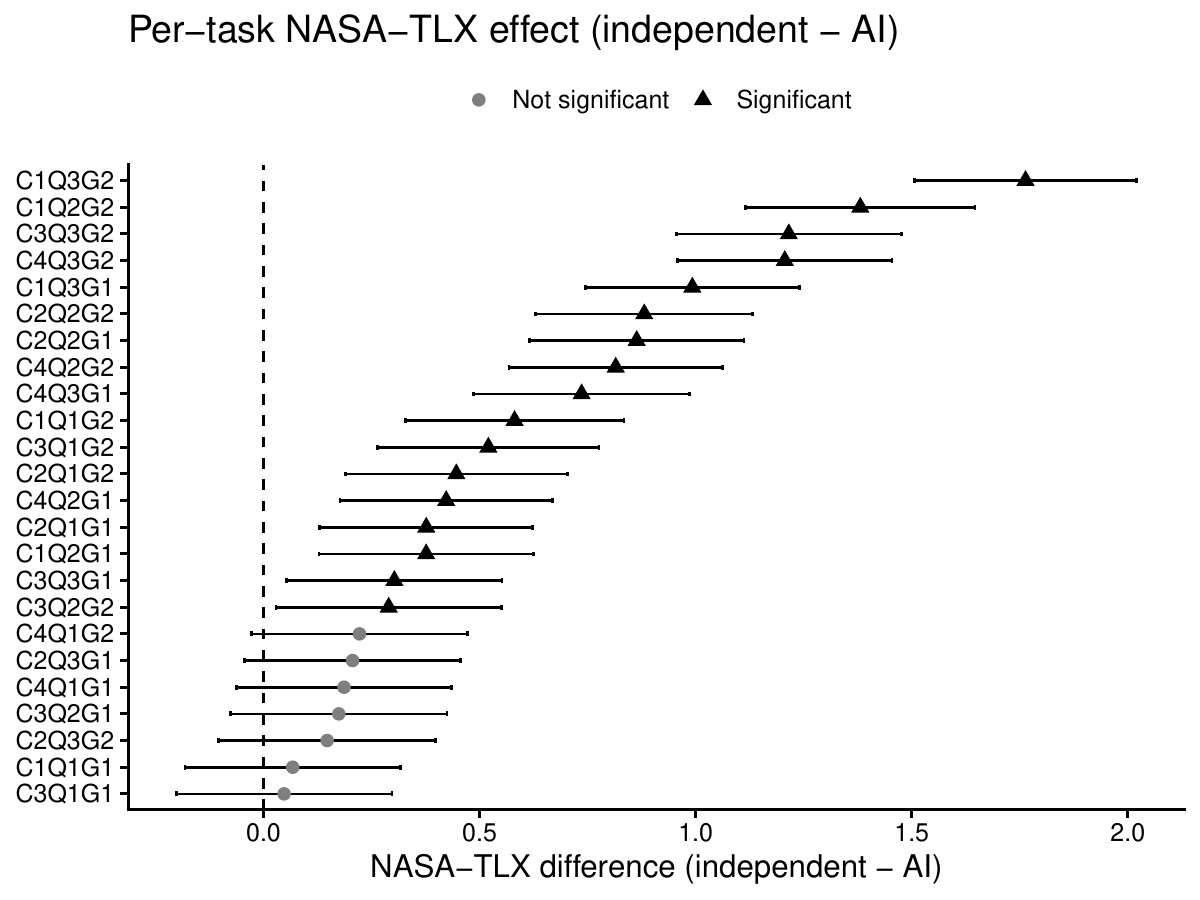}
    \end{subfigure}
    
    \caption{\textbf{AI-assisted completion time is faster for 6 tasks and slower for 5 tasks, while AI assistance lowers NASA-TLX for most tasks.} The difference between the two conditions (independent vs. AI-assisted) per task in terms of time (left) and NASA-TLX (right), ordered by effect size.}
    \label{fig:combined}
\end{figure}

Specifically, we found that AI assistance makes 6 tasks faster: writing a longer product review (C4Q2G2; $\beta=89.31$, $p<0.001$), editing a long text (C4Q3G2; $\beta=81.09$, $p<0.001$), coming up with 5 other words for ``enjoy'' (C1Q3G2; $\beta=56.66$, $p<0.001$), summarizing a long paragraph (C2Q2G2; $\beta=37.30$, $p<0.001$), adding 7 numbers (C3Q3G2; $\beta=31.31$, $p<0.01$), and summarizing a short paragraph (C2Q2G1; $\beta=22.83$, $p<0.05$). At the same time, we identified 5 tasks where AI assistance slows down task completion times: naming the color form mixing red, orange, yellow, green, and purple (C1Q1G2; $\beta=-40.55$, $p<0.001$), answering a simple trolley problem (C2Q3G1; $\beta=-34.73$, $p<0.01$), naming the color from mixing white and black (C1Q1G1: $\beta=-30.90$, $p<0.01$), naming one Olympic medalist (C1Q2G1; $\beta=-30.53$,  $p<0.01$), and the longer trolley problem (C2Q3G2: $\beta=-22.09$,  $p<0.05$); see Figure \ref{fig:combined}.

\subsection{Individual Traits} \label{appendix:study2_traits}

We investigate whether NFC and AIAS predict the magnitude of the speedup illusion. We quantify the magnitude of the speedup illusion as the difference between the predicted independent completion times and the predicted AI-assisted completion times (i.e. how much faster a participant believes external assistance can bring): $\hat{t}_H(\tau) - \hat{t}_A(\tau)$.

Using the prediction sample, we calculated the difference between the predicted independent time and predicted assisted time (where the source of assistance is either AI or another participant). We found that participants who are more averse to thinking (scoring lower on NFC) predicted a greater difference between the two predicted times ($\beta=-5.41$, $p<0.05$). This means that the more participants dislike having to think, the more susceptible they are to the speedup illusion, believing that using AI can be significantly faster than doing the task on their own. The finding is consistent with previous research indicating a stronger effect between perceived difficulty and subjective cognitive effort for individuals with low ability or conscientiousness \citep{yeo2008subjective}. On the other hand, the AIAS scale does not predict the magnitude of calibration error ($\beta=0.53$, $p=0.62$).

\subsection{Individual items from NASA-TLX} \label{appendix:study2_nasa_items}

\begin{table}[h]
\centering
\begin{tabular}{llcccc}
\hline
\textbf{Comparison} & \textbf{Item} & \textbf{Estimate} & \textbf{SE} & \textbf{z-ratio} & \textbf{p-value} \\
\hline

\multicolumn{6}{l}{\textit{Predicted Independent vs. Predicted AI}} \\
 & mental demand & 1.39 & 0.06 & 22.39 & $<.0001$ \\
 & hurried & 0.94 & 0.06 & 15.14 & $<.0001$ \\
 & successful & 0.32 & 0.06 & 5.42  & $<.0001$ \\
 & hard work & 1.37 & 0.07 & 20.68 & $<.0001$ \\
 & insecure  & 0.51 & 0.07 & 7.40  & $<.0001$ \\

\multicolumn{6}{l}{\textit{Predicted Independent vs. Predicted Another Participant}} \\
 & mental demand & 0.26  & 0.06 & 4.20  & $<.0001$ \\
 & hurried & 0.12  & 0.06 & 1.95  & 0.05 \\
 & successful & 0.05 & 0.06 & 0.76  & 0.45 \\
 & hard work & 0.26  & 0.07 & 3.94  & 0.0001 \\
 & insecure & -0.08 & 0.07 & -1.18 & 0.24 \\

\multicolumn{6}{l}{\textit{Actual vs. Predicted (Independent)}} \\
 & mental demand & -0.26 & 0.06 & -4.20 & $<.0001$ \\
 & hurried & -0.49 & 0.06 & -8.42 & $<.0001$ \\
  & successful & -0.09 & 0.06 & -1.58 & 0.11 \\
 & hard work & -0.26 & 0.06 & -4.19 & $<.0001$ \\
 & insecure & -0.39 & 0.06 & -6.95 & $<.0001$ \\

\multicolumn{6}{l}{\textit{Actual vs. Predicted (AI)}} \\
 & mental demand& 0.18  & 0.06 & 2.87  & 0.004 \\
 & hurried & 0.02 & 0.06 & 0.40  & 0.69 \\
 & successful & -0.22 & 0.06 & -3.75 & $<.0001$ \\
 & hard work & 0.13  & 0.07 & 1.97  & 0.05 \\
 & insecure & -0.20 & 0.06 & -3.42 & 0.0006 \\

\multicolumn{6}{l}{\textit{Actual Independent vs. Actual AI}} \\
 &mental demand& 0.94 & 0.07 & 13.49 & $<.0001$ \\
 & hurried & 0.36 & 0.07 & 4.80  & $<.0001$ \\
 & successful & 0.40 & 0.06 & 6.80  & $<.0001$ \\
 & hard work & 0.95 & 0.07 & 13.07 & $<.0001$ \\
 & insecure & 0.30 & 0.06 & 4.76  & $<.0001$ \\

\hline
\end{tabular}
\caption{Pairwise contrasts across prediction and completion conditions for the five items in NASA-TLX. }
\label{tab:nasa_tlx_contrasts}
\end{table}

Previous research (e.g. \citet{liu1994mental}) showed that the sub-scales of NASA-TLX captured different dimensions of effort. We perform an in-depth analysis of whether there are any item-level variations of the NASA-TLX scale and whether our general conclusions (using the average of the items) generalize. We find that for effort prediction, all items were predicted to be lower for the AI condition than the independent condition (see Table \ref{tab:nasa_tlx_contrasts}). However, when comparing the predicted effort for the independent condition with the another-participant condition, three items were no longer significant. The variation highlights that while participants predicted AI to be more helpful for every aspect that constitutes the NASA-TLX scale, when the assistance source is another participant, participants only found the effect to be present in item 1 (mental demand) and item 4 (hard work).

Even though no difference was found between the actual and predicted NASA-TLX average score when completing tasks with AI (see last section in Table \ref{tab:nasa_tlx_contrasts}), we identified three items with a significant difference: participants underestimated the rating required to complete a task with AI for item 1, item 3, and item 5. The significant differences highlight that item-wise variations exist within the scale, and even though the predicted and actual item values are both low for the AI condition, participants still predicted effort to be even lower than the actual experienced effort for these specific items.

\subsection{Robustness to prediction order} \label{appendix:prediction_order}
To ensure that the predicted time and effort are not biased as a result of the order of prediction, we randomized the prediction order where half of the participants predicted effort before predicting for time, and the other half predicted time before effort. To test whether the order has an effect on the prediction outcomes, we fitted the following model: $\text{predicted time} \sim \text{order} * \text{prediction target}  + (1 + \text{prediction target}\mid \text{participantID}) + (1 \mid \text{taskID})$. \\
We found no significant effect between the two orders ($\beta=6.25$, $p=0.08$) in this model. However, we found a significant effect where the predicted time for participants themselves was greater when participants predicted effort before time, compared with the predicted independent completion times when participants predicted time before effort ($\beta=9.43$, $p<0.05$). No significant difference was found for the assisted condition ($\beta=3.06$, $p=0.44$). Similarly, no significant difference was found between the predicted effort across the two orderings ($\beta=0.05$, $p=0.41$) for the independent condition ($\beta=0.09$, $p=0.19$) or the assisted condition ($\beta=0.01$, $p=0.89$). The results confirm that our conclusions are robust against the different prediction orders.

\subsection{How are people prompting AI?} \label{appendix:study2_behavior}

To understand the different ways that participants prompted the LLM, we provide an in-depth analysis of prompting behaviors and user-LLM interactions in the experiment. More than 81\% of the interactions are single-turn.  Fitting the model $\text{time} \sim \text{condition} * \text{copypaste} + \text{difficulty} + (1  \mid \text{participantID}) + (1 \mid \text{taskID})$, we found that copying and pasting significantly reduced completion time (for the AI condition) from 102.0 seconds to 66.2 seconds ($\beta=35.7$, $p<0.001$), but no significant difference between effort was found ($\beta=0.08$,  $p=0.07$). 

While copying the tasks directly as prompts was common, we also observed instances where participants wrote very different prompts even for the same task. For example, for the task on naming Olympic medalists, many participants had a specific category or name in mind and used AI as a recall tool (see Table \ref{tab:prompt_list} for examples). Some participants used the LLM to confirm or expand on their existing answers. For the trolley problems, where participants had to justify whether to divert the trolley, some participants asked AI to articulate the reasoning rather than making the decision (e.g., ``I have a situation where I can save 5 people by diverting a runaway troller. Nobody gets hurt so I want to do it. Give me a 3 sentence justification''). When using AI to summarize the paragraph, some participants chose to have AI expand on their own summary rather than delegating AI to do the task entirely, e.g., 
``OK, here's the story…And here's my sentence summarizing that, and I'm asking for your input on it: At dawn, as Mara departed, she decided to leave behind all memories of what life had led her to up to this point, and she stepped lightly and tenuously out into her brave new world.'' In these cases, the participants already expended the majority of the effort required to complete the task prior to receiving the AI answer.

Finally, we observed that some participants actively engaged with the model responses and asked questions in ways that best helped them solve the problem. For example, for the addition problem, one participant decomposed the problem into multiple two value addition problems, first asking ``what is 672+328'', then ``what is 1189+76'', and finally ``1265+59+56?''. For the water-bottle review-writing task, a participant provided clarification for the model response and added ``1 SENTENCE ONLY, DON'T USE ANY CONJOINT SENTENCE'' and ``DON'T USE ANY PUNCTUATION, KEEP IT TIGHTLY ONLY ONE SENTENCE'' as followup prompts. The wide range of behaviors showcase how differently participants prompted the model and the different degrees of cognitive engagement underlying the process.

\begin{table*}[t]
\small
\centering
\begin{tabular}{p{0.95\textwidth}}
\toprule
\textbf{Prompt} \\
\midrule
Is Simone Biles an Olympic medalist? \\

I would like to know the person who has the most gold medals in swimming, please. \\

Can you name who won the gold medal for female ice skating at the most recent Olympics?
 \\

Is jack Hughes an olympic athlete? \\

What was the name of the Olympic medalist who jumped off diving boards? I think he had brown hair.  \\

Name the male speed skater who won multiple medals at the 2026 winter Olympics
 \\

\bottomrule
\end{tabular}
\caption{Examples of participants' prompts to AI for the ``name an Olympic medalist'' task that were not directly copied.}
\label{tab:prompt_list}
\end{table*}

\section{Study 3 details} \label{appendix:study3}
\subsection{Statistical testing} \label{appendix:study3_stats}
\begin{figure}[t]
    \centering
    
    \begin{subfigure}[t]{0.48\columnwidth}
        \centering
        \includegraphics[width=\linewidth]{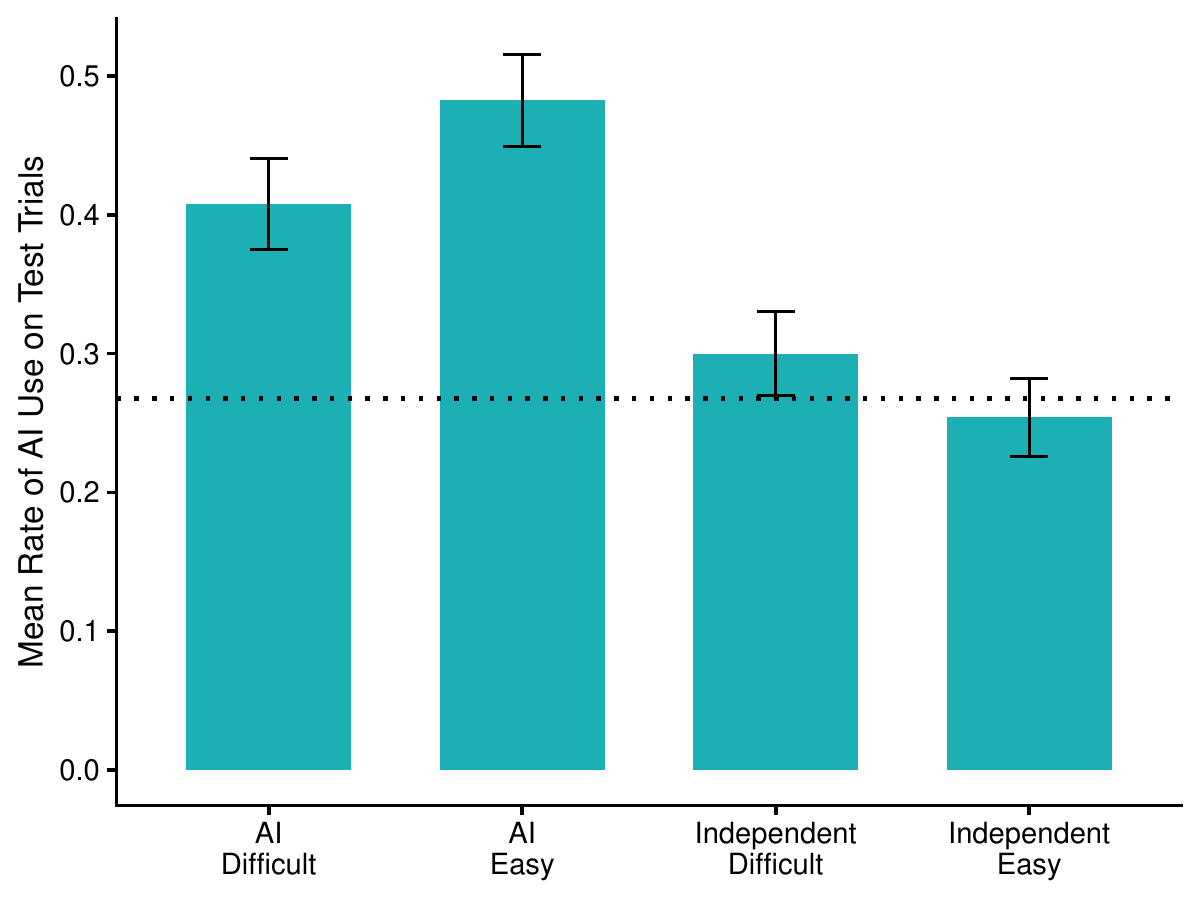}
    \end{subfigure}
    \hfill
    \begin{subfigure}[t]{0.48\columnwidth}
        \centering
        \includegraphics[width=\linewidth]{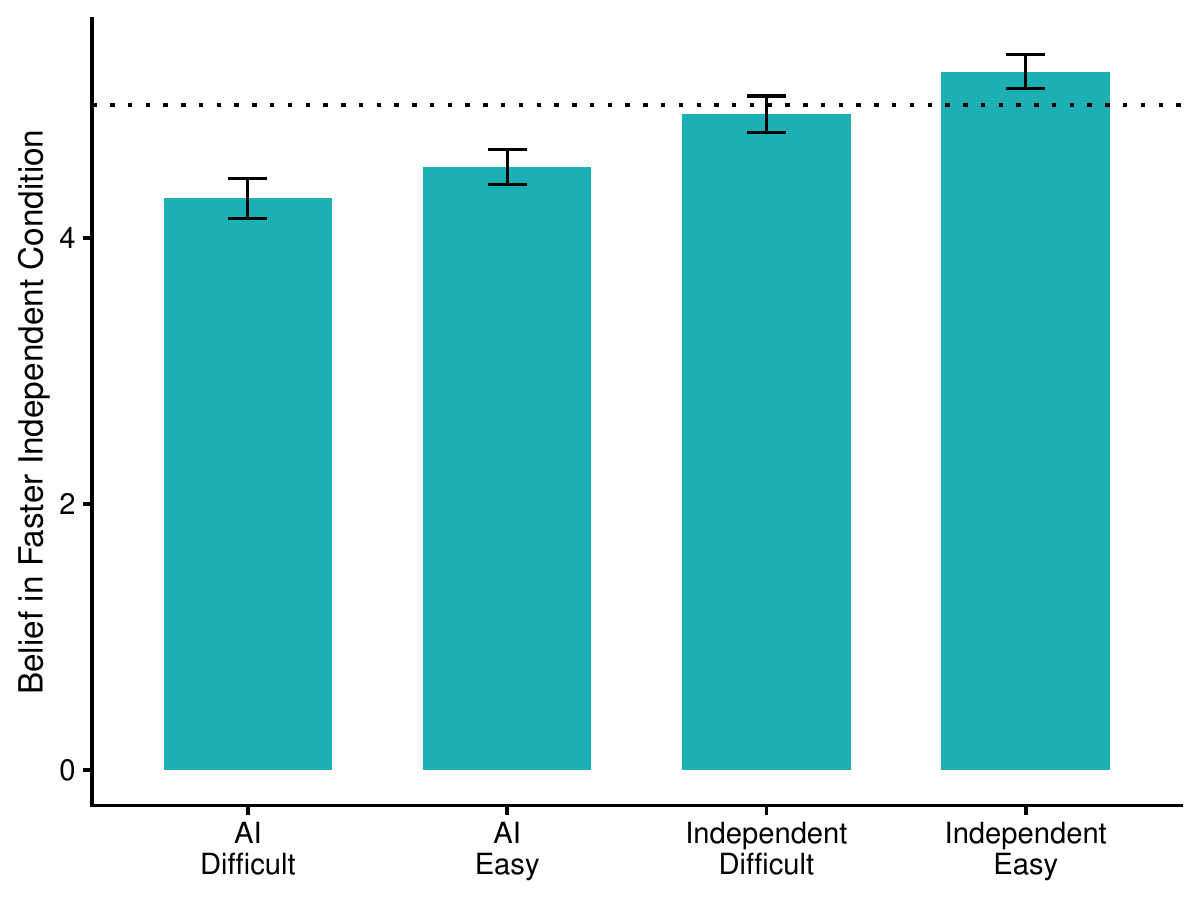}
    \end{subfigure}
    
    \caption{The subsequent rate of AI use on test trials (left) and the mean belief in faster independent completion (right) across conditions, compared to the control condition rate (shown in the dotted line). The subsequent rate of AI adoption on test trials is significantly greater than the control condition for participants who completed tasks (easy and difficult) with AI in the exposure phase, while the rate is not necessarily lower for participants who completed tasks independently in the exposure phase. Similarly, the belief in faster independent completion lowers compared to the control condition for participants who completed tasks with AI in the exposure phase.}
    \label{fig:study3_appendix}
\end{figure}

We fitted a mixed-effect logistic regression model $\text{AI use} \sim  \text{independent} + \text{NFC} + (1  \mid \text{participantID}) + (1 \mid \text{taskID})$ to predict AI use (where AI use is coded as whether participants prompted the model at least once for each of the last two tasks). The variable ``independent'' indicates whether participants were assigned into condition 3 or 4, where they had to complete tasks independently in the exposure phase, or condition 1 or 2, where they had to prompt an AI. To test the contrasts across the five conditions and compare the likelihood of using AI for the experimental conditions with the control condition, we replaced ``independent'' with ``condition'', where the latter variable is one of the five possible conditions in the study. Table \ref{tab:study3_detailed_stats} shows the pairwise contrasts across each pair of conditions, and Figure \ref{fig:study3_appendix} shows the average rate of subsequent AI use for each condition.

Both condition 1 (difficult, AI) and condition 2 (easy, AI) had a higher likelihood of predicting AI use than the control condition (condition 5); moreover, participants from condition 1 (difficult, AI) were more likely to use AI on subsequent tasks than participants from condition 4 (easy, independent); participants from condition 2 (easy, AI) are more likely to use AI than participants from condition 3 (difficult, independent) as well as participants from condition 4 (easy, independent). 

When the dependent variable is the outcome variables on the 7-point likert scale, we simplified the prediction model to $\text{variable} \sim  \text{independent} + \text{NFC} + (1  \mid \text{participantID})$, where the variable is either perceived difficulty, time calibration, effort calibration, or confidence, rated on the 7-point scale.

\begin{table}[h]
\centering
\begin{tabular}{lrrrr}
\hline
\textbf{Contrast} & \textbf{Estimate} & \textbf{SE} & \textbf{z-ratio} & \textbf{p-value} \\
\hline
condition5 $-$ condition1  & 0.87 & 0.30 & 2.91 & 0.04 \\
condition5 $-$ condition2  & -1.33 & 0.30 & 4.37 & 0.0001 \\
condition5 $-$ condition3  & 0.11 & 0.30 & 0.36 & 1 \\
condition5 $-$ condition4  &  -0.10 & 0.30 &  -0.32 & 1 \\
condition1 $-$ condition2  & -0.46 & 0.29 & -1.59 & 1\\
condition1 $-$ condition3  &  0.76 & 0.30 &  2.53 & 0.12 \\
condition1 $-$ condition4  &  0.96 & 0.31 &  3.16 & 0.02\\
condition2 $-$ condition3  &  1.22 & 0.31 &  3.99 & 0.0007 \\
condition2 $-$ condition4  &  1.42 & 0.31 &  4.59 & $<.0001$ \\
condition3 $-$ condition4  &  0.21 & 0.31 &  0.68 & 1\\
\hline
\end{tabular}
\caption{Pairwise contrasts between conditions for the mixed-effects logistic regression model. Condition 5 is the control condition. In condition 1, participants completed two difficult task variants with AI; in condition 2, participants completed two easy task variants with AI; in condition 3, participants completed two difficult task variants independently; in condition 4, participants completed two easy task variants independently.}
\label{tab:study3_detailed_stats}
\end{table}

\subsection{Reasons for using vs. not using AI} \label{appendix:study3_qualitative}
To better understand why and how participants made the decision to complete the last two tasks independently or using AI, we included an open-ended question where participants shared why they chose to complete the task the way they did. Across the responses, time savings were a huge factor in shaping people's decision to either use or not use AI. Both participants who chose to use AI and participants who did not reported that it was because they believed that their selected completion mode would require less time, confirming that the decision to offload can be understood as a resource-rational decision to conserve time.

For people who chose to use AI, other reasons besides time savings included effort savings, confidence and trust in AI outputs, ability constraints, curiosity in the AI generation, and general inertia. Similarly, reasons for not using AI included time savings, effort savings, people's preference for independent thought, and a lack of trust in AI. Table \ref{tab:ai_use_reasons} shows the clusters of reasons for both groups from qualitative analysis and example quotes for each cluster. Even though people had other reasons for using or not using AI (e.g. personal preference), we found that time savings and effort savings are two recurring justifications, highlighting that people's different calibrations directly led to the different choice to use or not use AI --- while participants who believed that they could complete the tasks faster did not use AI, participants who believed otherwise decided to use AI --- showing that time and effort calibration directly drove behavior.

\begin{table}[ht]
\centering
\small
\begin{tabular}{p{0.22\textwidth} p{0.68\textwidth}}
\hline
\textbf{Reasons} & \textbf{Example Responses} \\
\hline

\multicolumn{2}{l}{\textbf{Reasons for Using AI}} \\
\hline

Time savings &
``time was the main factor, whichever would [be] quicker'' \newline
``I used AI because it would be quicker to do so than to write out the 3 sentences by myself.'' \\

Effort savings &
``The second task took more effort, therefore I chose to use AI.'' \newline
``I chose to use AI because it would be easier.'' \\

Confidence/trust in AI &
``In a study, I became anxious and worried I would somehow get the answer wrong, so using AI helped.'' \newline
``The first one it was faster to not use it the second one u used it to make sure i was right.'' \\

Ability constraints &
``I could not think of a word for the last task. the others i knew the answers to.'' \newline
``I used AI for the first task because I'm not great at math.'' \\

Curiosity &
``I had an answer, but also wanted to see what the AI would answer.'' \newline
``it was more fun to use AI and see what it would say, and also it was less work and faster.'' \\

Availability/inertia &
``I used AI because it was there.'' \newline
``I chose to complete the two tasks using AI because I was already using the chatbot.'' \newline
``I decided to use AI because I was not sure of the name of an Olympic gold medalist off the top of my head. And I used it the second time as well because it just seemed like the easier option and didn't require me to think too much about it.'' \\

\hline
\multicolumn{2}{l}{\textbf{Reasons for Not Using AI}} \\
\hline

Time savings &
``I don't like using AI, and feel like it would have taken me longer to do so.'' \newline
``I thought I could answer them quicker.'' \\

Effort savings &
``I chose to complete all independently as I found it easier than using AI, I honestly feel like am cheating when am using AI. I did not use it for any of the tasks. '' \newline
``I chose to complete the tasks independently because it was faster and easier than using AI.'' \\

Preferred thinking &
``I wanted to complete them independently to use my own brain rather than rely on AI.'' \newline
``I am older than dirt, and can do math in my head which is faster than typing a prompt.'' \\

Trust &
``I trusted myself more than ai and it was faster.'' \newline
``Because AI sucks and its not worth the resources...'' \\

Confidence &
``If I am confident of an answer there is no point in using AI.'' \\

Unfamiliarity with AI &
``I am not very familiar with using AI for more than just finding information.'' \\

\hline
\end{tabular}
\caption{Representative reasons participants gave for using AI versus completing tasks independently.}
\label{tab:ai_use_reasons}
\end{table}

\subsection{Outcome variables predict AI use}

In this section, we investigate the correlation between perceived difficulty, time calibration, effort calibration, and confidence and AI use on the last two tasks. We found that participants who perceived the last two tasks as difficult were more likely to use AI on the tasks ($\beta=0.27$, $p<0.01$). Time calibration negatively predicts AI use: the more strongly a participant believes a task would be faster to complete independently, the less likely they are to use AI ($\beta=-0.50$, $p<0.001$). Conversely, the more strongly a participant believed a task would require more effort to complete independently, the more likely they are to use AI on the task ($\beta=0.41$, $p<0.001$). Finally, the more confident a participant is in their response, the less likely they would use AI ($\beta=-0.09$, $p<0.001$). The correlations confirm that calibrations about time and effort are directly related to behaviors surrounding AI use.

\subsection{First vs. Second Task} \label{appendix:study3_first_second}

\begin{figure}[t]
    \centering
    
    \begin{subfigure}[t]{0.48\columnwidth}
        \centering
        \includegraphics[width=\linewidth]{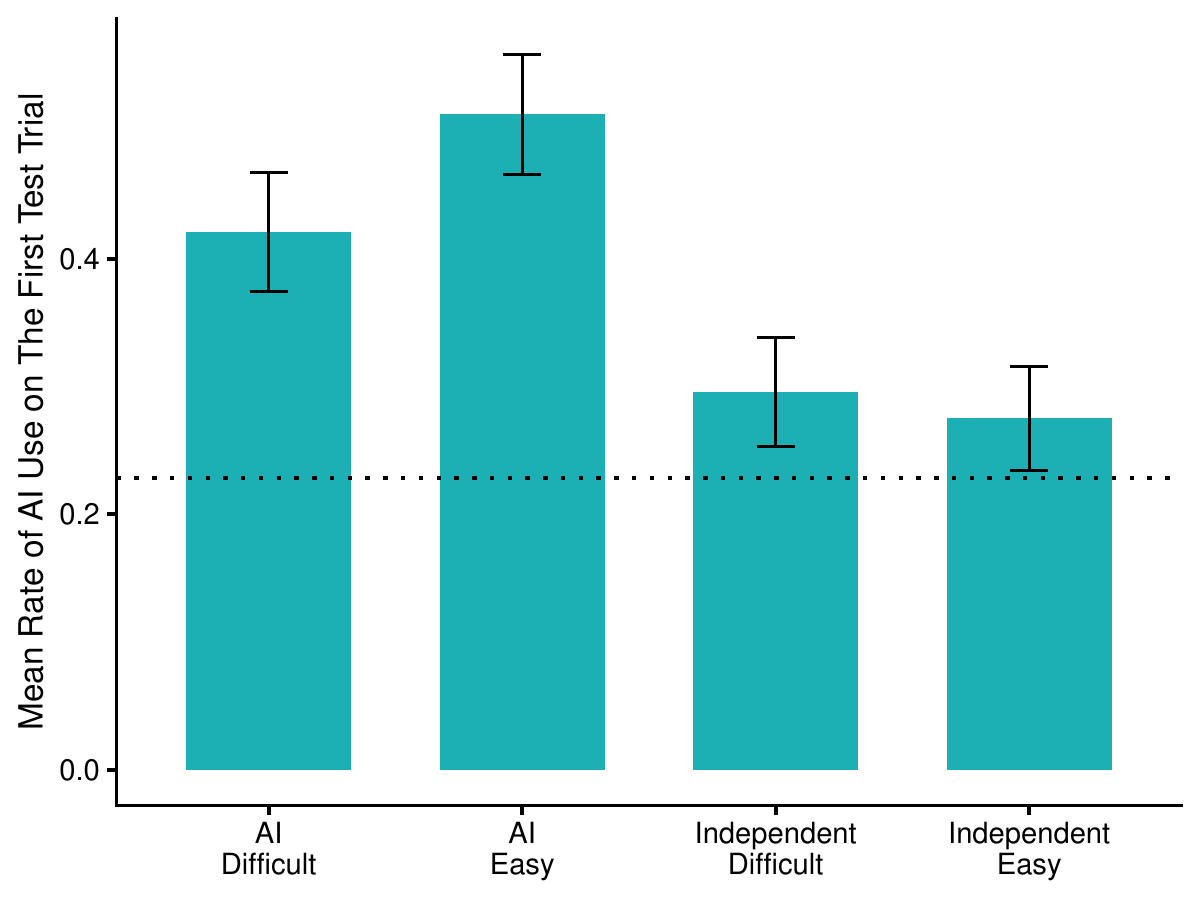}
    \end{subfigure}
    \hfill
    \begin{subfigure}[t]{0.48\columnwidth}
        \centering
        \includegraphics[width=\linewidth]{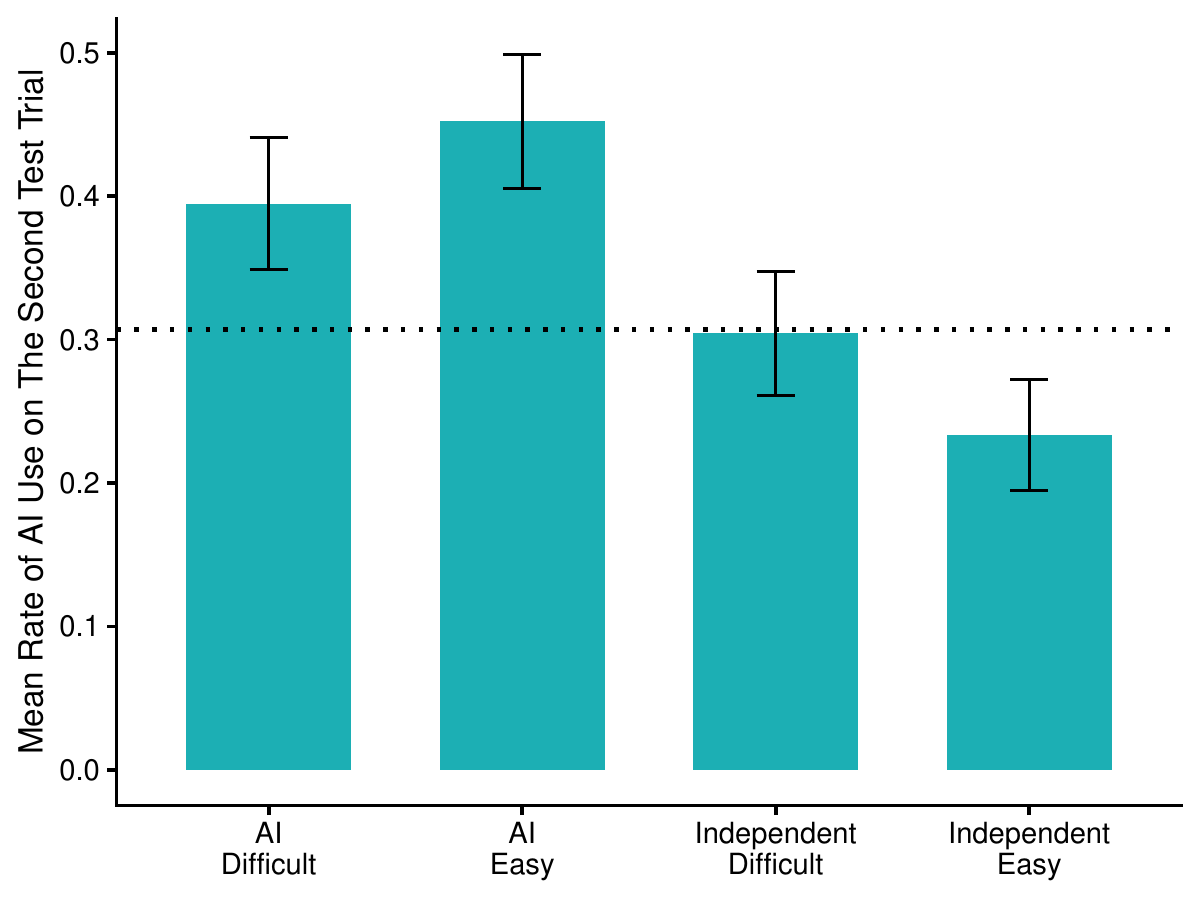}
    \end{subfigure}
    
    \caption{\textbf{Our conclusions are robust whether only including data from the first or second test trial}. The rate of AI use for conditions where participants were exposed to using AI is consistently higher than conditions where participants completed tasks independently, regardless of the first task (left) or the second task (right).}
    \label{fig:study3_details}
\end{figure}

In the main analysis, we looked at the rate of AI use on both tasks in the test trial, but behaviors might differ because of the order of the tasks. Here we confirm that whether only looking at AI use on the first task in the test phase ($\beta=0.42$, $p<0.001$) or the second task ($\beta=0.38$, $p<0.001$), the likelihood of AI use is greater for participants who are exposed to completing tasks with AI (see Figure \ref{fig:study3_details}), confirming the robustness of our findings.

\end{appendices}

\FloatBarrier

\bibliography{sn-bibliography}


\begin{thebibliography}{60}
\ifx \bisbn   \undefined \def \bisbn  #1{ISBN #1}\fi
\ifx \binits  \undefined \def \binits#1{#1}\fi
\ifx \bauthor  \undefined \def \bauthor#1{#1}\fi
\ifx \batitle  \undefined \def \batitle#1{#1}\fi
\ifx \bjtitle  \undefined \def \bjtitle#1{#1}\fi
\ifx \bvolume  \undefined \def \bvolume#1{\textbf{#1}}\fi
\ifx \byear  \undefined \def \byear#1{#1}\fi
\ifx \bissue  \undefined \def \bissue#1{#1}\fi
\ifx \bfpage  \undefined \def \bfpage#1{#1}\fi
\ifx \blpage  \undefined \def \blpage #1{#1}\fi
\ifx \burl  \undefined \def \burl#1{\textsf{#1}}\fi
\ifx \doiurl  \undefined \def \doiurl#1{\url{https://doi.org/#1}}\fi
\ifx \betal  \undefined \def \betal{\textit{et al.}}\fi
\ifx \binstitute  \undefined \def \binstitute#1{#1}\fi
\ifx \binstitutionaled  \undefined \def \binstitutionaled#1{#1}\fi
\ifx \bctitle  \undefined \def \bctitle#1{#1}\fi
\ifx \beditor  \undefined \def \beditor#1{#1}\fi
\ifx \bpublisher  \undefined \def \bpublisher#1{#1}\fi
\ifx \bbtitle  \undefined \def \bbtitle#1{#1}\fi
\ifx \bedition  \undefined \def \bedition#1{#1}\fi
\ifx \bseriesno  \undefined \def \bseriesno#1{#1}\fi
\ifx \blocation  \undefined \def \blocation#1{#1}\fi
\ifx \bsertitle  \undefined \def \bsertitle#1{#1}\fi
\ifx \bsnm \undefined \def \bsnm#1{#1}\fi
\ifx \bsuffix \undefined \def \bsuffix#1{#1}\fi
\ifx \bparticle \undefined \def \bparticle#1{#1}\fi
\ifx \barticle \undefined \def \barticle#1{#1}\fi
\bibcommenthead
\ifx \bconfdate \undefined \def \bconfdate #1{#1}\fi
\ifx \botherref \undefined \def \botherref #1{#1}\fi
\ifx \url \undefined \def \url#1{\textsf{#1}}\fi
\ifx \bchapter \undefined \def \bchapter#1{#1}\fi
\ifx \bbook \undefined \def \bbook#1{#1}\fi
\ifx \bcomment \undefined \def \bcomment#1{#1}\fi
\ifx \oauthor \undefined \def \oauthor#1{#1}\fi
\ifx \citeauthoryear \undefined \def \citeauthoryear#1{#1}\fi
\ifx \endbibitem  \undefined \def \endbibitem {}\fi
\ifx \bconflocation  \undefined \def \bconflocation#1{#1}\fi
\ifx \arxivurl  \undefined \def \arxivurl#1{\textsf{#1}}\fi
\csname PreBibitemsHook\endcsname

\bibitem[\protect\citeauthoryear{Collins et~al.}{2024}]{collins2024building}
\begin{barticle}
\bauthor{\bsnm{Collins}, \binits{K.M.}},
\bauthor{\bsnm{Sucholutsky}, \binits{I.}},
\bauthor{\bsnm{Bhatt}, \binits{U.}},
\bauthor{\bsnm{Chandra}, \binits{K.}},
\bauthor{\bsnm{Wong}, \binits{L.}},
\bauthor{\bsnm{Lee}, \binits{M.}},
\bauthor{\bsnm{Zhang}, \binits{C.E.}},
\bauthor{\bsnm{Zhi-Xuan}, \binits{T.}},
\bauthor{\bsnm{Ho}, \binits{M.}},
\bauthor{\bsnm{Mansinghka}, \binits{V.}},
\bauthor{\bsnm{Weller}, \binits{A.}},
\bauthor{\bsnm{Tenenbaum}, \binits{J.B.}},
\bauthor{\bsnm{Griffiths}, \binits{T.L.o.}}:
\batitle{Building machines that learn and think with people}.
\bjtitle{Nature human behaviour}
\bvolume{8}(\bissue{10}),
\bfpage{1851}--\blpage{1863}
(\byear{2024})
\end{barticle}
\endbibitem

\bibitem[\protect\citeauthoryear{Zao-Sanders}{2025}]{zao2025people}
\begin{botherref}
\oauthor{\bsnm{Zao-Sanders}, \binits{M.}}:
How people are really using gen {AI} in 2025.
Harvard Business Review
\textbf{9}
(2025)
\end{botherref}
\endbibitem

\bibitem[\protect\citeauthoryear{Cheng et~al.}{2026}]{cheng2026sycophantic}
\begin{barticle}
\bauthor{\bsnm{Cheng}, \binits{M.}},
\bauthor{\bsnm{Lee}, \binits{C.}},
\bauthor{\bsnm{Khadpe}, \binits{P.}},
\bauthor{\bsnm{Yu}, \binits{S.}},
\bauthor{\bsnm{Han}, \binits{D.}},
\bauthor{\bsnm{Jurafsky}, \binits{D.}}:
\batitle{Sycophantic {AI} decreases prosocial intentions and promotes dependence}.
\bjtitle{Science}
\bvolume{391}(\bissue{6792}),
\bfpage{8352}
(\byear{2026})
\end{barticle}
\endbibitem

\bibitem[\protect\citeauthoryear{Handa et~al.}{2025}]{handa2025economic}
\begin{botherref}
\oauthor{\bsnm{Handa}, \binits{K.}},
\oauthor{\bsnm{Tamkin}, \binits{A.}},
\oauthor{\bsnm{McCain}, \binits{M.}},
\oauthor{\bsnm{Huang}, \binits{S.}},
\oauthor{\bsnm{Durmus}, \binits{E.}},
\oauthor{\bsnm{Heck}, \binits{S.}},
\oauthor{\bsnm{Mueller}, \binits{J.}},
\oauthor{\bsnm{Hong}, \binits{J.}},
\oauthor{\bsnm{Ritchie}, \binits{S.}},
\oauthor{\bsnm{Belonax}, \binits{T.}}, et al.:
Which economic tasks are performed with ai? evidence from millions of claude conversations.
arXiv preprint arXiv:2503.04761
(2025)
\end{botherref}
\endbibitem

\bibitem[\protect\citeauthoryear{Wang et~al.}{2025}]{wang2025ai}
\begin{botherref}
\oauthor{\bsnm{Wang}, \binits{Z.Z.}},
\oauthor{\bsnm{Shao}, \binits{Y.}},
\oauthor{\bsnm{Shaikh}, \binits{O.}},
\oauthor{\bsnm{Fried}, \binits{D.}},
\oauthor{\bsnm{Neubig}, \binits{G.}},
\oauthor{\bsnm{Yang}, \binits{D.}}:
How do {AI} agents do human work? comparing {AI} and human workflows across diverse occupations.
arXiv preprint arXiv:2510.22780
(2025)
\end{botherref}
\endbibitem

\bibitem[\protect\citeauthoryear{Tamkin and McCrory}{2025}]{tamkinmccrory2025productivity}
\begin{botherref}
\oauthor{\bsnm{Tamkin}, \binits{A.}},
\oauthor{\bsnm{McCrory}, \binits{P.}}:
Estimating {AI} Productivity Gains from Claude Conversations.
\url{https://www.anthropic.com/research/estimating-productivity-gains}
\end{botherref}
\endbibitem

\bibitem[\protect\citeauthoryear{Appel et~al.}{2026}]{anthropic2026aeiv4}
\begin{botherref}
\oauthor{\bsnm{Appel}, \binits{R.}},
\oauthor{\bsnm{Massenkoff}, \binits{M.}},
\oauthor{\bsnm{McCrory}, \binits{P.}},
\oauthor{\bsnm{McCain}, \binits{M.}},
\oauthor{\bsnm{Heller}, \binits{R.}},
\oauthor{\bsnm{Neylon}, \binits{T.}},
\oauthor{\bsnm{Tamkin}, \binits{A.}}:
Anthropic Economic Index Report: Economic Primitives.
\url{https://www.anthropic.com/research/anthropic-economic-index-january-2026-report}
\end{botherref}
\endbibitem

\bibitem[\protect\citeauthoryear{Xiong et~al.}{2024}]{xiong2024search}
\begin{botherref}
\oauthor{\bsnm{Xiong}, \binits{H.}},
\oauthor{\bsnm{Bian}, \binits{J.}},
\oauthor{\bsnm{Li}, \binits{Y.}},
\oauthor{\bsnm{Li}, \binits{X.}},
\oauthor{\bsnm{Du}, \binits{M.}},
\oauthor{\bsnm{Wang}, \binits{S.}},
\oauthor{\bsnm{Yin}, \binits{D.}},
\oauthor{\bsnm{Helal}, \binits{S.}}:
When search engine services meet large language models: visions and challenges.
IEEE Transactions on Services Computing
(2024)
\end{botherref}
\endbibitem

\bibitem[\protect\citeauthoryear{Hooper}{2025}]{hooper2025cognitive}
\begin{bchapter}
\bauthor{\bsnm{Hooper}, \binits{V.J.}}:
\bctitle{Cognitive offloading and the reshaping of human thought: The subtle influence of artificial intelligence}.
In: \bbtitle{Colloquia, Academic Journal of Culture and Thought},
vol. \bseriesno{12},
pp. \bfpage{01}--\blpage{14}
(\byear{2025})
\end{bchapter}
\endbibitem

\bibitem[\protect\citeauthoryear{Stadler et~al.}{2024}]{stadler2024cognitive}
\begin{barticle}
\bauthor{\bsnm{Stadler}, \binits{M.}},
\bauthor{\bsnm{Bannert}, \binits{M.}},
\bauthor{\bsnm{Sailer}, \binits{M.}}:
\batitle{Cognitive ease at a cost: Llms reduce mental effort but compromise depth in student scientific inquiry}.
\bjtitle{Computers in Human Behavior}
\bvolume{160},
\bfpage{108386}
(\byear{2024})
\end{barticle}
\endbibitem

\bibitem[\protect\citeauthoryear{Dhillon et~al.}{2024}]{dhillon2024shaping}
\begin{bchapter}
\bauthor{\bsnm{Dhillon}, \binits{P.S.}},
\bauthor{\bsnm{Molaei}, \binits{S.}},
\bauthor{\bsnm{Li}, \binits{J.}},
\bauthor{\bsnm{Golub}, \binits{M.}},
\bauthor{\bsnm{Zheng}, \binits{S.}},
\bauthor{\bsnm{Robert}, \binits{L.P.}}:
\bctitle{Shaping human-{AI} collaboration: Varied scaffolding levels in co-writing with language models}.
In: \bbtitle{Proceedings of the 2024 CHI Conference on Human Factors in Computing Systems},
pp. \bfpage{1}--\blpage{18}
(\byear{2024})
\end{bchapter}
\endbibitem

\bibitem[\protect\citeauthoryear{Becker et~al.}{2025}]{becker2025measuring}
\begin{botherref}
\oauthor{\bsnm{Becker}, \binits{J.}},
\oauthor{\bsnm{Rush}, \binits{N.}},
\oauthor{\bsnm{Barnes}, \binits{E.}},
\oauthor{\bsnm{Rein}, \binits{D.}}:
Measuring the Impact of Early-2025 {AI} on Experienced Open-Source Developer Productivity.
\url{https://metr.org/blog/2025-07-10-early-2025-AI-experienced-os-dev-study/}
(2025)
\end{botherref}
\endbibitem

\bibitem[\protect\citeauthoryear{Risko and Gilbert}{2016}]{risko2016cognitive}
\begin{barticle}
\bauthor{\bsnm{Risko}, \binits{E.F.}},
\bauthor{\bsnm{Gilbert}, \binits{S.J.}}:
\batitle{Cognitive offloading}.
\bjtitle{Trends in cognitive sciences}
\bvolume{20}(\bissue{9}),
\bfpage{676}--\blpage{688}
(\byear{2016})
\end{barticle}
\endbibitem

\bibitem[\protect\citeauthoryear{Gerlich}{2025}]{gerlich2025ai}
\begin{barticle}
\bauthor{\bsnm{Gerlich}, \binits{M.}}:
\batitle{{AI} tools in society: Impacts on cognitive offloading and the future of critical thinking}.
\bjtitle{Societies}
\bvolume{15}(\bissue{1}),
\bfpage{6}
(\byear{2025})
\end{barticle}
\endbibitem

\bibitem[\protect\citeauthoryear{Lee et~al.}{2025}]{lee2025impact}
\begin{bchapter}
\bauthor{\bsnm{Lee}, \binits{H.-P.}},
\bauthor{\bsnm{Sarkar}, \binits{A.}},
\bauthor{\bsnm{Tankelevitch}, \binits{L.}},
\bauthor{\bsnm{Drosos}, \binits{I.}},
\bauthor{\bsnm{Rintel}, \binits{S.}},
\bauthor{\bsnm{Banks}, \binits{R.}},
\bauthor{\bsnm{Wilson}, \binits{N.}}:
\bctitle{The impact of generative {AI} on critical thinking: Self-reported reductions in cognitive effort and confidence effects from a survey of knowledge workers}.
In: \bbtitle{Proceedings of the 2025 CHI Conference on Human Factors in Computing Systems},
pp. \bfpage{1}--\blpage{22}
(\byear{2025})
\end{bchapter}
\endbibitem

\bibitem[\protect\citeauthoryear{Liu et~al.}{2026}]{liu2026ai}
\begin{botherref}
\oauthor{\bsnm{Liu}, \binits{G.}},
\oauthor{\bsnm{Christian}, \binits{B.}},
\oauthor{\bsnm{Dumbalska}, \binits{T.}},
\oauthor{\bsnm{Bakker}, \binits{M.A.}},
\oauthor{\bsnm{Dubey}, \binits{R.}}:
{AI} assistance reduces persistence and hurts independent performance.
arXiv preprint arXiv:2604.04721
(2026)
\end{botherref}
\endbibitem

\bibitem[\protect\citeauthoryear{Barcaui}{2025}]{barcaui2025chatgpt}
\begin{barticle}
\bauthor{\bsnm{Barcaui}, \binits{A.}}:
\batitle{{ChatGPT} as a cognitive crutch: Evidence from a randomized controlled trial on knowledge retention}.
\bjtitle{Social Sciences \& Humanities Open}
\bvolume{12},
\bfpage{102287}
(\byear{2025})
\end{barticle}
\endbibitem

\bibitem[\protect\citeauthoryear{Shen and Tamkin}{2026}]{shen2026ai}
\begin{botherref}
\oauthor{\bsnm{Shen}, \binits{J.H.}},
\oauthor{\bsnm{Tamkin}, \binits{A.}}:
How {AI} impacts skill formation.
arXiv preprint arXiv:2601.20245
(2026)
\end{botherref}
\endbibitem

\bibitem[\protect\citeauthoryear{Shelby et~al.}{2025}]{shelby2025taxonomy}
\begin{botherref}
\oauthor{\bsnm{Shelby}, \binits{R.}},
\oauthor{\bsnm{Diaz}, \binits{F.}},
\oauthor{\bsnm{Prabhakaran}, \binits{V.}}:
Taxonomy of user needs and actions.
arXiv preprint arXiv:2510.06124
(2025)
\end{botherref}
\endbibitem

\bibitem[\protect\citeauthoryear{Festinger}{1962}]{festinger1962cognitive}
\begin{barticle}
\bauthor{\bsnm{Festinger}, \binits{L.}}:
\batitle{Cognitive dissonance}.
\bjtitle{Scientific American}
\bvolume{207}(\bissue{4}),
\bfpage{93}--\blpage{106}
(\byear{1962})
\end{barticle}
\endbibitem

\bibitem[\protect\citeauthoryear{Hart and Staveland}{1988}]{Hart_1986}
\begin{botherref}
\oauthor{\bsnm{Hart}, \binits{S.G.}},
\oauthor{\bsnm{Staveland}, \binits{L.E.}}:
Development of {NASA-TLX} (task load index): Results of empirical and theoretical research.
vol. 52,
pp. 139--183.
Elsevier
(1988)
\end{botherref}
\endbibitem

\bibitem[\protect\citeauthoryear{Koriat}{2015}]{koriat2015metacognition}
\begin{barticle}
\bauthor{\bsnm{Koriat}, \binits{A.}}:
\batitle{Metacognition: Decision making processes in self-monitoring and self-regulation}.
\bjtitle{The Wiley Blackwell handbook of judgment and decision making}
\bvolume{2},
\bfpage{356}--\blpage{379}
(\byear{2015})
\end{barticle}
\endbibitem

\bibitem[\protect\citeauthoryear{Yeung and Summerfield}{2012}]{yeung2012metacognition}
\begin{barticle}
\bauthor{\bsnm{Yeung}, \binits{N.}},
\bauthor{\bsnm{Summerfield}, \binits{C.}}:
\batitle{Metacognition in human decision-making: confidence and error monitoring}.
\bjtitle{Philosophical Transactions of the Royal Society B: Biological Sciences}
\bvolume{367}(\bissue{1594}),
\bfpage{1310}--\blpage{1321}
(\byear{2012})
\end{barticle}
\endbibitem

\bibitem[\protect\citeauthoryear{Fleming and Daw}{2017}]{fleming2017self}
\begin{barticle}
\bauthor{\bsnm{Fleming}, \binits{S.M.}},
\bauthor{\bsnm{Daw}, \binits{N.D.}}:
\batitle{Self-evaluation of decision-making: A general bayesian framework for metacognitive computation.}
\bjtitle{Psychological review}
\bvolume{124}(\bissue{1}),
\bfpage{91}
(\byear{2017})
\end{barticle}
\endbibitem

\bibitem[\protect\citeauthoryear{Dubey et~al.}{2026}]{dubey2021aha}
\begin{barticle}
\bauthor{\bsnm{Dubey}, \binits{R.}},
\bauthor{\bsnm{Ho}, \binits{M.}},
\bauthor{\bsnm{Mehta}, \binits{H.}},
\bauthor{\bsnm{Griffiths}, \binits{T.L.}}:
\batitle{Aha! moments correspond to metacognitive prediction errors}.
\bjtitle{Cognition}
\bvolume{274},
\bfpage{106537}
(\byear{2026})
\doiurl{10.1016/j.cognition.2026.106537}
\end{barticle}
\endbibitem

\bibitem[\protect\citeauthoryear{Cacioppo and Petty}{1982}]{cacioppo1982need}
\begin{barticle}
\bauthor{\bsnm{Cacioppo}, \binits{J.T.}},
\bauthor{\bsnm{Petty}, \binits{R.E.}}:
\batitle{The need for cognition.}
\bjtitle{Journal of personality and social psychology}
\bvolume{42}(\bissue{1}),
\bfpage{116}
(\byear{1982})
\end{barticle}
\endbibitem

\bibitem[\protect\citeauthoryear{Leimeister}{2010}]{leimeister2010collective}
\begin{barticle}
\bauthor{\bsnm{Leimeister}, \binits{J.M.}}:
\batitle{Collective intelligence}.
\bjtitle{Business \& Information Systems Engineering}
\bvolume{2}(\bissue{4}),
\bfpage{245}--\blpage{248}
(\byear{2010})
\end{barticle}
\endbibitem

\bibitem[\protect\citeauthoryear{Fan et~al.}{2023}]{fan2023drawing}
\begin{barticle}
\bauthor{\bsnm{Fan}, \binits{J.E.}},
\bauthor{\bsnm{Bainbridge}, \binits{W.A.}},
\bauthor{\bsnm{Chamberlain}, \binits{R.}},
\bauthor{\bsnm{Wammes}, \binits{J.D.}}:
\batitle{Drawing as a versatile cognitive tool}.
\bjtitle{Nature Reviews Psychology}
\bvolume{2}(\bissue{9}),
\bfpage{556}--\blpage{568}
(\byear{2023})
\end{barticle}
\endbibitem

\bibitem[\protect\citeauthoryear{Lieder and Griffiths}{2020}]{lieder2020resource}
\begin{barticle}
\bauthor{\bsnm{Lieder}, \binits{F.}},
\bauthor{\bsnm{Griffiths}, \binits{T.L.}}:
\batitle{Resource-rational analysis: Understanding human cognition as the optimal use of limited computational resources}.
\bjtitle{Behavioral and brain sciences}
\bvolume{43},
\bfpage{1}
(\byear{2020})
\end{barticle}
\endbibitem

\bibitem[\protect\citeauthoryear{Griffiths et~al.}{2019}]{griffiths2019doing}
\begin{barticle}
\bauthor{\bsnm{Griffiths}, \binits{T.L.}},
\bauthor{\bsnm{Callaway}, \binits{F.}},
\bauthor{\bsnm{Chang}, \binits{M.B.}},
\bauthor{\bsnm{Grant}, \binits{E.}},
\bauthor{\bsnm{Krueger}, \binits{P.M.}},
\bauthor{\bsnm{Lieder}, \binits{F.}}:
\batitle{Doing more with less: meta-reasoning and meta-learning in humans and machines}.
\bjtitle{Current Opinion in Behavioral Sciences}
\bvolume{29},
\bfpage{24}--\blpage{30}
(\byear{2019})
\end{barticle}
\endbibitem

\bibitem[\protect\citeauthoryear{Griffiths}{2020}]{griffiths2020understanding}
\begin{barticle}
\bauthor{\bsnm{Griffiths}, \binits{T.L.}}:
\batitle{Understanding human intelligence through human limitations}.
\bjtitle{Trends in Cognitive Sciences}
\bvolume{24}(\bissue{11}),
\bfpage{873}--\blpage{883}
(\byear{2020})
\end{barticle}
\endbibitem

\bibitem[\protect\citeauthoryear{Dunn and Risko}{2016}]{dunn2016toward}
\begin{barticle}
\bauthor{\bsnm{Dunn}, \binits{T.L.}},
\bauthor{\bsnm{Risko}, \binits{E.F.}}:
\batitle{Toward a metacognitive account of cognitive offloading}.
\bjtitle{Cognitive Science}
\bvolume{40}(\bissue{5}),
\bfpage{1080}--\blpage{1127}
(\byear{2016})
\end{barticle}
\endbibitem

\bibitem[\protect\citeauthoryear{Wahn et~al.}{2023}]{wahn2023offloading}
\begin{barticle}
\bauthor{\bsnm{Wahn}, \binits{B.}},
\bauthor{\bsnm{Schmitz}, \binits{L.}},
\bauthor{\bsnm{Gerster}, \binits{F.N.}},
\bauthor{\bsnm{Weiss}, \binits{M.}}:
\batitle{Offloading under cognitive load: Humans are willing to offload parts of an attentionally demanding task to an algorithm}.
\bjtitle{Plos one}
\bvolume{18}(\bissue{5}),
\bfpage{0286102}
(\byear{2023})
\end{barticle}
\endbibitem

\bibitem[\protect\citeauthoryear{Messeri and Crockett}{2024}]{messeri2024artificial}
\begin{barticle}
\bauthor{\bsnm{Messeri}, \binits{L.}},
\bauthor{\bsnm{Crockett}, \binits{M.J.}}:
\batitle{Artificial intelligence and illusions of understanding in scientific research}.
\bjtitle{Nature}
\bvolume{627}(\bissue{8002}),
\bfpage{49}--\blpage{58}
(\byear{2024})
\end{barticle}
\endbibitem

\bibitem[\protect\citeauthoryear{Kim et~al.}{2026}]{kim2026llm}
\begin{botherref}
\oauthor{\bsnm{Kim}, \binits{H.}},
\oauthor{\bsnm{Yu}, \binits{H.}},
\oauthor{\bsnm{Yi}, \binits{H.}}:
The {LLM} fallacy: Misattribution in {AI-}assisted cognitive workflows.
arXiv preprint arXiv:2604.14807
(2026)
\end{botherref}
\endbibitem

\bibitem[\protect\citeauthoryear{Fraisse}{1984}]{fraisse1984perception}
\begin{barticle}
\bauthor{\bsnm{Fraisse}, \binits{P.}}:
\batitle{Perception and estimation of time}.
\bjtitle{Annual review of psychology}
\bvolume{35}(\bissue{1}),
\bfpage{1}--\blpage{37}
(\byear{1984})
\end{barticle}
\endbibitem

\bibitem[\protect\citeauthoryear{Fredrickson and Kahneman}{1993}]{fredrickson1993duration}
\begin{barticle}
\bauthor{\bsnm{Fredrickson}, \binits{B.L.}},
\bauthor{\bsnm{Kahneman}, \binits{D.}}:
\batitle{Duration neglect in retrospective evaluations of affective episodes.}
\bjtitle{Journal of personality and social psychology}
\bvolume{65}(\bissue{1}),
\bfpage{45}
(\byear{1993})
\end{barticle}
\endbibitem

\bibitem[\protect\citeauthoryear{Zauberman et~al.}{2009}]{zauberman2009discounting}
\begin{barticle}
\bauthor{\bsnm{Zauberman}, \binits{G.}},
\bauthor{\bsnm{Kim}, \binits{B.K.}},
\bauthor{\bsnm{Malkoc}, \binits{S.A.}},
\bauthor{\bsnm{Bettman}, \binits{J.R.}}:
\batitle{Discounting time and time discounting: Subjective time perception and intertemporal preferences}.
\bjtitle{Journal of Marketing Research}
\bvolume{46}(\bissue{4}),
\bfpage{543}--\blpage{556}
(\byear{2009})
\end{barticle}
\endbibitem

\bibitem[\protect\citeauthoryear{Liu and Wickens}{1994}]{liu1994mental}
\begin{barticle}
\bauthor{\bsnm{Liu}, \binits{Y.}},
\bauthor{\bsnm{Wickens}, \binits{C.D.}}:
\batitle{Mental workload and cognitive task automaticity: an evaluation of subjective and time estimation metrics}.
\bjtitle{Ergonomics}
\bvolume{37}(\bissue{11}),
\bfpage{1843}--\blpage{1854}
(\byear{1994})
\end{barticle}
\endbibitem

\bibitem[\protect\citeauthoryear{Matthews and Meck}{2016}]{matthews2016temporal}
\begin{barticle}
\bauthor{\bsnm{Matthews}, \binits{W.J.}},
\bauthor{\bsnm{Meck}, \binits{W.H.}}:
\batitle{Temporal cognition: Connecting subjective time to perception, attention, and memory.}
\bjtitle{Psychological bulletin}
\bvolume{142}(\bissue{8}),
\bfpage{865}
(\byear{2016})
\end{barticle}
\endbibitem

\bibitem[\protect\citeauthoryear{Brickman}{1971}]{brickman1971hedonic}
\begin{botherref}
\oauthor{\bsnm{Brickman}, \binits{P.}}:
Hedonic relativism and planning the good society.
Adaptation level theory,
287--301
(1971)
\end{botherref}
\endbibitem

\bibitem[\protect\citeauthoryear{Brickman et~al.}{1978}]{brickman1978lottery}
\begin{barticle}
\bauthor{\bsnm{Brickman}, \binits{P.}},
\bauthor{\bsnm{Coates}, \binits{D.}},
\bauthor{\bsnm{Janoff-Bulman}, \binits{R.}}:
\batitle{Lottery winners and accident victims: Is happiness relative?}
\bjtitle{Journal of personality and social psychology}
\bvolume{36}(\bissue{8}),
\bfpage{917}
(\byear{1978})
\end{barticle}
\endbibitem

\bibitem[\protect\citeauthoryear{Frederick and Loewenstein}{1999}]{frederick199916}
\begin{barticle}
\bauthor{\bsnm{Frederick}, \binits{S.}},
\bauthor{\bsnm{Loewenstein}, \binits{G.}}:
\batitle{16 hedonic adaptation}.
\bjtitle{Well-Being The foundations of hedonic psychology}
\bvolume{63},
\bfpage{302}--\blpage{329}
(\byear{1999})
\end{barticle}
\endbibitem

\bibitem[\protect\citeauthoryear{Oktar et~al.}{2026}]{oktar2025identifying}
\begin{barticle}
\bauthor{\bsnm{Oktar}, \binits{K.}},
\bauthor{\bsnm{Collins}, \binits{K.M.}},
\bauthor{\bsnm{Hern\'{a}ndez-Orallo}, \binits{J.}},
\bauthor{\bsnm{Coyle}, \binits{D.}},
\bauthor{\bsnm{Cave}, \binits{S.}},
\bauthor{\bsnm{Weller}, \binits{A.}},
\bauthor{\bsnm{Sucholutsky}, \binits{I.}}:
\batitle{Identifying, evaluating, and mitigating risks of {AI} thought partnerships}.
\bjtitle{ACM {AI} Lett.}
(\byear{2026})
\doiurl{10.1145/3803024}
\end{barticle}
\endbibitem

\bibitem[\protect\citeauthoryear{Ahn}{2025}]{ahn2025preserving}
\begin{barticle}
\bauthor{\bsnm{Ahn}, \binits{S.}}:
\batitle{Preserving critical thinking in the age of large language models: The paradox of cognitive load and efficiency}.
\bjtitle{The Korean Journal of Medicine}
\bvolume{100}(\bissue{5}),
\bfpage{197}--\blpage{200}
(\byear{2025})
\end{barticle}
\endbibitem

\bibitem[\protect\citeauthoryear{Hofman et~al.}{2023}]{hofman2023steroids}
\begin{botherref}
\oauthor{\bsnm{Hofman}, \binits{J.}},
\oauthor{\bsnm{Goldstein}, \binits{D.G.}},
\oauthor{\bsnm{Rothschild}, \binits{D.}}:
A sports analogy for understanding different ways to use ai.
Harvard Business Review
(2023)
\end{botherref}
\endbibitem

\bibitem[\protect\citeauthoryear{Jose et~al.}{2025}]{jose2025outsourcing}
\begin{barticle}
\bauthor{\bsnm{Jose}, \binits{B.}},
\bauthor{\bsnm{Joseph}, \binits{D.}},
\bauthor{\bsnm{Mohan}, \binits{V.}},
\bauthor{\bsnm{Alexander}, \binits{E.}},
\bauthor{\bsnm{Varghese}, \binits{S.K.}},
\bauthor{\bsnm{Roy}, \binits{A.}}:
\batitle{Outsourcing cognition: the psychological costs of {AI-}era convenience}.
\bjtitle{Frontiers in Psychology}
\bvolume{16},
\bfpage{1645237}
(\byear{2025})
\end{barticle}
\endbibitem

\bibitem[\protect\citeauthoryear{Collins et~al.}{2025}]{collins2025revisiting}
\begin{botherref}
\oauthor{\bsnm{Collins}, \binits{K.M.}},
\oauthor{\bsnm{Bhatt}, \binits{U.}},
\oauthor{\bsnm{Sucholutsky}, \binits{I.}}:
Revisiting rogers' paradox in the context of human-{AI} interaction.
arXiv preprint arXiv:2501.10476
(2025)
\end{botherref}
\endbibitem

\bibitem[\protect\citeauthoryear{Shaw and Nave}{2026}]{shaw2026thinking}
\begin{botherref}
\oauthor{\bsnm{Shaw}, \binits{S.}},
\oauthor{\bsnm{Nave}, \binits{G.}}:
Thinking—fast, slow, and artificial: How {AI} is reshaping human reasoning and the rise of cognitive surrender. psyarxiv.
Preprint at https://osf. io/preprints/psyarxiv/yk25n\_v1
(2026)
\end{botherref}
\endbibitem

\bibitem[\protect\citeauthoryear{Ibrahim et~al.}{2025}]{ibrahim2025measuring}
\begin{botherref}
\oauthor{\bsnm{Ibrahim}, \binits{L.}},
\oauthor{\bsnm{Collins}, \binits{K.M.}},
\oauthor{\bsnm{Kim}, \binits{S.S.Y.}},
\oauthor{\bsnm{Reuel}, \binits{A.}},
\oauthor{\bsnm{Lamparth}, \binits{M.}},
\oauthor{\bsnm{Feng}, \binits{K.}},
\oauthor{\bsnm{Ahmad}, \binits{L.}},
\oauthor{\bsnm{Soni}, \binits{P.}},
\oauthor{\bsnm{Kattan}, \binits{A.E.}},
\oauthor{\bsnm{Stein}, \binits{M.}},
\oauthor{\bsnm{Swaroop}, \binits{S.}},
\oauthor{\bsnm{Sucholutsky}, \binits{I.}},
\oauthor{\bsnm{Strait}, \binits{A.}},
\oauthor{\bsnm{Liao}, \binits{Q.V.}},
\oauthor{\bsnm{Bhatt}, \binits{U.}}:
Measuring and mitigating overreliance is necessary for building human-compatible {AI}.
arXiv preprint arXiv:2509.08010
(2025)
\end{botherref}
\endbibitem

\bibitem[\protect\citeauthoryear{Sturgeon et~al.}{2025}]{sturgeon2025humanagencybench}
\begin{botherref}
\oauthor{\bsnm{Sturgeon}, \binits{B.}},
\oauthor{\bsnm{Samuelson}, \binits{D.}},
\oauthor{\bsnm{Haimes}, \binits{J.}},
\oauthor{\bsnm{Anthis}, \binits{J.R.}}:
Humanagencybench: Scalable evaluation of human agency support in {AI} assistants.
arXiv preprint arXiv:2509.08494
(2025)
\end{botherref}
\endbibitem

\bibitem[\protect\citeauthoryear{Elizondo et~al.}{2024}]{elizondo2024self}
\begin{barticle}
\bauthor{\bsnm{Elizondo}, \binits{K.}},
\bauthor{\bsnm{Valenzuela}, \binits{R.}},
\bauthor{\bsnm{Pestana}, \binits{J.V.}},
\bauthor{\bsnm{Codina}, \binits{N.}}:
\batitle{Self-regulation and procrastination in college students: A tale of motivation, strategy, and perseverance}.
\bjtitle{Psychology in the Schools}
\bvolume{61}(\bissue{3}),
\bfpage{887}--\blpage{902}
(\byear{2024})
\end{barticle}
\endbibitem

\bibitem[\protect\citeauthoryear{Bansal et~al.}{2019}]{bansal2019beyond}
\begin{bchapter}
\bauthor{\bsnm{Bansal}, \binits{G.}},
\bauthor{\bsnm{Nushi}, \binits{B.}},
\bauthor{\bsnm{Kamar}, \binits{E.}},
\bauthor{\bsnm{Lasecki}, \binits{W.S.}},
\bauthor{\bsnm{Weld}, \binits{D.S.}},
\bauthor{\bsnm{Horvitz}, \binits{E.}}:
\bctitle{Beyond accuracy: The role of mental models in human-{AI} team performance}.
In: \bbtitle{Proceedings of the {AAAI} Conference on Human Computation and Crowdsourcing},
vol. \bseriesno{7},
pp. \bfpage{2}--\blpage{11}
(\byear{2019})
\end{bchapter}
\endbibitem

\bibitem[\protect\citeauthoryear{Kelly et~al.}{2023}]{kelly2023capturing}
\begin{bchapter}
\bauthor{\bsnm{Kelly}, \binits{M.}},
\bauthor{\bsnm{Kumar}, \binits{A.}},
\bauthor{\bsnm{Smyth}, \binits{P.}},
\bauthor{\bsnm{Steyvers}, \binits{M.}}:
\bctitle{Capturing humans’ mental models of ai: An item response theory approach}.
In: \bbtitle{Proceedings of the 2023 ACM Conference on Fairness, Accountability, and Transparency},
pp. \bfpage{1723}--\blpage{1734}
(\byear{2023})
\end{bchapter}
\endbibitem

\bibitem[\protect\citeauthoryear{Collins et~al.}{2024}]{collins2024modulating}
\begin{botherref}
\oauthor{\bsnm{Collins}, \binits{K.M.}},
\oauthor{\bsnm{Chen}, \binits{V.}},
\oauthor{\bsnm{Sucholutsky}, \binits{I.}},
\oauthor{\bsnm{Kirk}, \binits{H.R.}},
\oauthor{\bsnm{Sadek}, \binits{M.}},
\oauthor{\bsnm{Sargeant}, \binits{H.}},
\oauthor{\bsnm{Talwalkar}, \binits{A.}},
\oauthor{\bsnm{Weller}, \binits{A.}},
\oauthor{\bsnm{Bhatt}, \binits{U.}}:
Modulating language model experiences through frictions.
arXiv preprint arXiv:2407.12804
(2024)
\end{botherref}
\endbibitem

\bibitem[\protect\citeauthoryear{Steele}{2020}]{steele2020perception}
\begin{botherref}
\oauthor{\bsnm{Steele}, \binits{J.}}:
What is (perception of) effort? Objective and subjective effort during attempted task performance.
PsyArXiv
(2020).
\doiurl{10.31234/osf.io/kbyhm} .
\url{osf.io/preprints/psyarxiv/kbyhm_v1}
\end{botherref}
\endbibitem

\bibitem[\protect\citeauthoryear{Grassini}{2023}]{grassini2023development}
\begin{barticle}
\bauthor{\bsnm{Grassini}, \binits{S.}}:
\batitle{Development and validation of the {AI} attitude scale (aias-4): a brief measure of general attitude toward artificial intelligence}.
\bjtitle{Frontiers in psychology}
\bvolume{14},
\bfpage{1191628}
(\byear{2023})
\end{barticle}
\endbibitem

\bibitem[\protect\citeauthoryear{Yurt and Kasarci}{2024}]{yurt2024questionnaire}
\begin{botherref}
\oauthor{\bsnm{Yurt}, \binits{E.}},
\oauthor{\bsnm{Kasarci}, \binits{I.}}:
A questionnaire of artificial intelligence use motives: A contribution to investigating the connection between {AI} and motivation.
International Journal of Technology in Education
\textbf{7}(2)
(2024)
\end{botherref}
\endbibitem

\bibitem[\protect\citeauthoryear{Al{\'o}s-Ferrer et~al.}{2016}]{alos2016inertia}
\begin{barticle}
\bauthor{\bsnm{Al{\'o}s-Ferrer}, \binits{C.}},
\bauthor{\bsnm{H{\"u}gelsch{\"a}fer}, \binits{S.}},
\bauthor{\bsnm{Li}, \binits{J.}}:
\batitle{Inertia and decision making}.
\bjtitle{Frontiers in psychology}
\bvolume{7},
\bfpage{169}
(\byear{2016})
\end{barticle}
\endbibitem

\bibitem[\protect\citeauthoryear{Yeo and Neal}{2008}]{yeo2008subjective}
\begin{barticle}
\bauthor{\bsnm{Yeo}, \binits{G.}},
\bauthor{\bsnm{Neal}, \binits{A.}}:
\batitle{Subjective cognitive effort: A model of states, traits, and time.}
\bjtitle{Journal of Applied Psychology}
\bvolume{93}(\bissue{3}),
\bfpage{617}
(\byear{2008})
\end{barticle}
\endbibitem

\end{thebibliography}
\end{document}